\documentclass[aps,prd,twocolumn,superscriptaddress,amssymb,eqsecnum,showpacs,showkeyes,secnumarabic,graphics,floatfix,nofootinbib,tightenlines,longbibliography]{revtex4-1}

\usepackage{graphicx}
\usepackage{bm}
\usepackage{cancel}
\usepackage{dcolumn}
\usepackage{amsmath}
\usepackage{hhline}
\usepackage[utf8]{inputenc}
\usepackage[breaklinks=true,colorlinks=true,
linkcolor=blue,urlcolor=blue,citecolor=cyan,
bookmarks=true,bookmarksopenlevel=2]{hyperref}

\def \beq  {\begin{equation}}
\def \eeq  {\end{equation}}
\def \ber  {\begin{eqnarray}}
\def \eer  {\end{eqnarray}}

\begin{document}
\newcommand{\newc}{\newcommand}

\newc{\be}{\begin{equation}}
\newc{\ee}{\end{equation}}
\newc{\ba}{\begin{eqnarray}}
\newc{\ea}{\end{eqnarray}}
\newc{\bea}{\begin{eqnarray*}}
\newc{\eea}{\end{eqnarray*}}
\newc{\D}{\partial}
\newc{\ie}{{\it i.e.} }
\newc{\eg}{{\it e.g.} }
\newc{\etc}{{\it etc.} }
\newc{\etal}{{\it et al.}}
\newc{\lcdm}{$\Lambda$CDM}
\newcommand{\nn}{\nonumber}
\newc{\ra}{\Rightarrow}
\title{A Comparison of Thawing and Freezing Dark Energy Parametrizations}
\author{G. Pantazis}\email{ph06767@cc.uoi.gr}
\affiliation{Department of Physics, University of Ioannina, GR-45110, Ioannina, Greece}
\author{S. Nesseris}\email{savvas.nesseris@csic.es}
\affiliation{Instituto de F\'isica Te\'orica UAM-CSIC, Universidad Auton\'oma de Madrid, Cantoblanco, 28049 Madrid, Spain}
\author{L. Perivolaropoulos}\email{leandros@uoi.gr}
\affiliation{Department of Physics, University of Ioannina, GR-45110, Ioannina, Greece}

\date {\today}

\begin{abstract}
Dark energy equation of state $w(z)$ parametrizations with two parameters and given monotonicity are generically either convex or concave functions. This makes them suitable for fitting either freezing or thawing quintessence models but not both simultaneously. Fitting a dataset based on a freezing model with  an unsuitable (concave when increasing) $w(z)$ parametrization (like CPL)  can lead to significant misleading features like crossing of the phantom divide line, incorrect $w(z=0)$, incorrect slope \etc that are not present in the underlying cosmological model. To demonstrate this fact we generate scattered cosmological data both at the level of $w(z)$ and the luminosity distance $D_L(z)$ based on either thawing or freezing quintessence models and fit  them using parametrizations of convex and of concave type. We then compare statistically significant features of the best fit $w(z)$ with actual features of the underlying model. We thus verify that  the use of unsuitable parametrizations can lead to misleading conclusions. In order to avoid these problems it is important to either use both convex and concave parametrizations and select the one with the best $\chi^2$ or use principal component analysis thus splitting the redshift range into independent bins. In the latter case however, significant information about the slope of $w(z)$ at high redshifts is lost. Finally, we propose a new family of parametrizations (nCPL) $w(z)=w_0+w_a (\frac{z}{1+z})^n$ which generalizes the CPL and interpolates between thawing and freezing parametrizations as the parameter $n$ increases to values larger than 1.
\end{abstract}
\maketitle

\section{Introduction}
\label{sec:Introduction}
The simplest cosmological model that is consistent with current cosmological observations is the \lcdm \ model \cite{Peebles:2002gy, Padmanabhan:2002ji, Carroll:2000fy}. According to this model, the observed accelerating expansion of the universe is attributed to the repulsive gravitational force of a cosmological constant which may be described as a perfect fluid (dark energy \cite{Tsujikawa:2010sc, Sami:2009jx, Frieman:2008sn, Alcaniz:2006ay}) with constant energy density $\rho$ and negative pressure $p$  with equation of state parameter
\be w=\frac{p}{\rho}=-1. \label{wdef} \ee

The cosmological observations that are consistent with \lcdm \ include type Ia supernovae \cite{{GonzalezGaitan:2011rg},{Sullivan:2011kv}}, baryon acoustic oscillations (BAO) \citep{Aubourg:2014yra}, anisotropies in the Cosmic Microwave Background (CMB) \cite{Komatsu:2008hk, Ade:2015xua, Ade:2015rim} and large scale structure \cite{Reid:2012sw, White:2014naa, Xu:2013tsa, Nesseris:2007pa, Linder:2005in, Basilakos:2014yda, Basilakos:2012uu, Avsajanishvili:2015elt}. These observations allow for significant deviations from \lcdm \ (\eg \cite{Joyce:2014kja}) which are usually parametrized through two parameter parametrizations of the form $w(z,w_0,w_a)$ where $w_0$, $w_a$ are parameters that estimate the deviation of the dark energy equation of state parameter from the \lcdm \ value $w=-1$. In the context of the plethora of possible parametrizations, the following questions arise

\begin{itemize}
\item[1.] Are there generally acceptable cosmological data that are inconsistent with \lcdm \ in a statistically significant manner?
\item[2.]  Is there a dark energy parametrization which fits the data better than \lcdm \ in a consistent and statistically significant manner?
\item[3.] If there is, what is the class of models it corresponds to?
\end{itemize}

At present the answer to the first question is negative despite the dramatic increase of accuracy of cosmological data \cite{Ade:2015xua, Array:2015xqh, Bull:2015stt, Dawson:2015wdb, Abbott:2015swa}. In view of this fact and the simplicity of \lcdm \ there has not been significant investigation in the direction of the second question even though the answer to this question is independent of the answer to the first question.

The standard parametrization used to compare with the fit of \lcdm \ is the Chevallier-Polarski-Linder (CPL) \cite{{Chevallier:2000qy},{Linder:2002et}} which is based on a linear  expansion with respect to the scale factor around its present value $a=1$ and is of the form
\be w(z)=w_0+w_a(1-a)=w_0 + w_a \frac{z}{1+z}, \label{cpl} \ee
where $z$ is the redshift corresponding to the scale factor $a$.

In the context of the CPL parametrization and a few others considered in the literature there appears to be no significant advantage over \lcdm \ \cite{Crittenden:2007yy}. However, there is currently no systematic strategy developed towards constructing parametrizations specially designed so that they remain simple and natural while at the same time fit the data better than \lcdm. Such parametrizations should also be designed to represent efficiently classes of physical models.

The simplest physical class of models which is an alternative to \lcdm \ is quintessence \cite{Tsujikawa:2013fta, Zlatev:1998tr, Carroll:1998zi, Caldwell:2005tm, Sahni:2002kh, Barreiro:1999zs, Chiba:2012cb, Thakur:2012rp, Wetterich:1987fm} where the role of the dark energy fluid is played by a minimally coupled to gravity scalar field with dynamics determined by a specially designed scalar field potential. This class of models reduces to \lcdm \ for a constant potential. It also shares the fine tuning problems of \lcdm \ since the energy density of the scalar field at present is required to be unnaturally low ($\rho_\text{DE}\simeq 10^{-47}GeV^4$) which corresponds to a scalar field mass $m_\phi \sim H_0 \simeq 10^{-33} eV$. However its dynamical degrees of freedom have the potential to make the recent domination of dark energy density over matter more natural.

Quintessence models may be classified in two broad categories \cite{Caldwell:2005tm, Linder:2007wa, Huterer:2006mv, Linder:2006sv, Werner:2005pi}: Thawing models \cite{Gupta:2014uea, Chiba:2009sj, Scherrer:2007pu} and freezing (or tracking) models \cite{Sahlen:2006dn, Schimd:2006pa, Chiba:2005tj, Scherrer:2005je}. In the first class the scalar field is frozen at early times due to the Hubble cosmic friction term while at late times when the Hubble parameter decreases, friction becomes subdominant. The dynamics due to the scalar field potential takes over while a non-zero kinetic term develops, thus increasing the equation of state parameter of the scalar field dark energy in accordance to the equation
\be w\equiv \frac{p_\phi}{\rho_\phi}= \frac{\dot{\phi}^2 /2-V(\phi)}{\dot{\phi}^2 /2+V(\phi)}, \label{eos} \ee
where $V(\phi)$ is the scalar field potential and $p_\phi$, $\rho_\phi$ are the pressure and energy density of the scalar field. Thus in thawing models, at early times  $w\simeq -1$, since the kinetic energy term is negligible and $w>-1$ at late times when the kinetic term starts to develop. In thawing models the function $w(z)$ is generically a monotonic convex decreasing function which reaches asymptotically at high redshifts $z$ the value $-1$.

In the second class of quintessence models (freezing-tracker models), the potential is steep enough at early times so that a kinetic term develops but the scalar field gradually slows down as the potential becomes shallower at late times. For example, for inverse power law potentials of the form $V(\phi)=M^{4+p} \phi^{-p} (p>0)$ the equation of state of the scalar field is constant during the matter era ($w\sim 0$) and decreases towards $w=-1$ at late times. Thus, in freezing models the function $w(z)$ is a generically monotonic convex  increasing function which asymptotes towards $w=-1$ at late times (low $z$). Due to the asymptotic approach of the value $w=-1$  at late times, we have $w(z)\sim z^n$  ($n>1$) at low $z$ for freezing models and therefore there is no linear term in the $w(z)$ expansion.

We therefore have two distinct types of behavior for $w(z)$ which are motivated by classes of physical models: decreasing convex function with $w(z=0)>-1$  (thawing behavior) and increasing concave function with $w(z=0)\simeq -1$  (freezing behavior).

\begin{figure}[!t]
\centering
\vspace{0cm}\rotatebox{0}{\vspace{0cm}\hspace{0cm}\resizebox{0.49\textwidth}{!}{\includegraphics{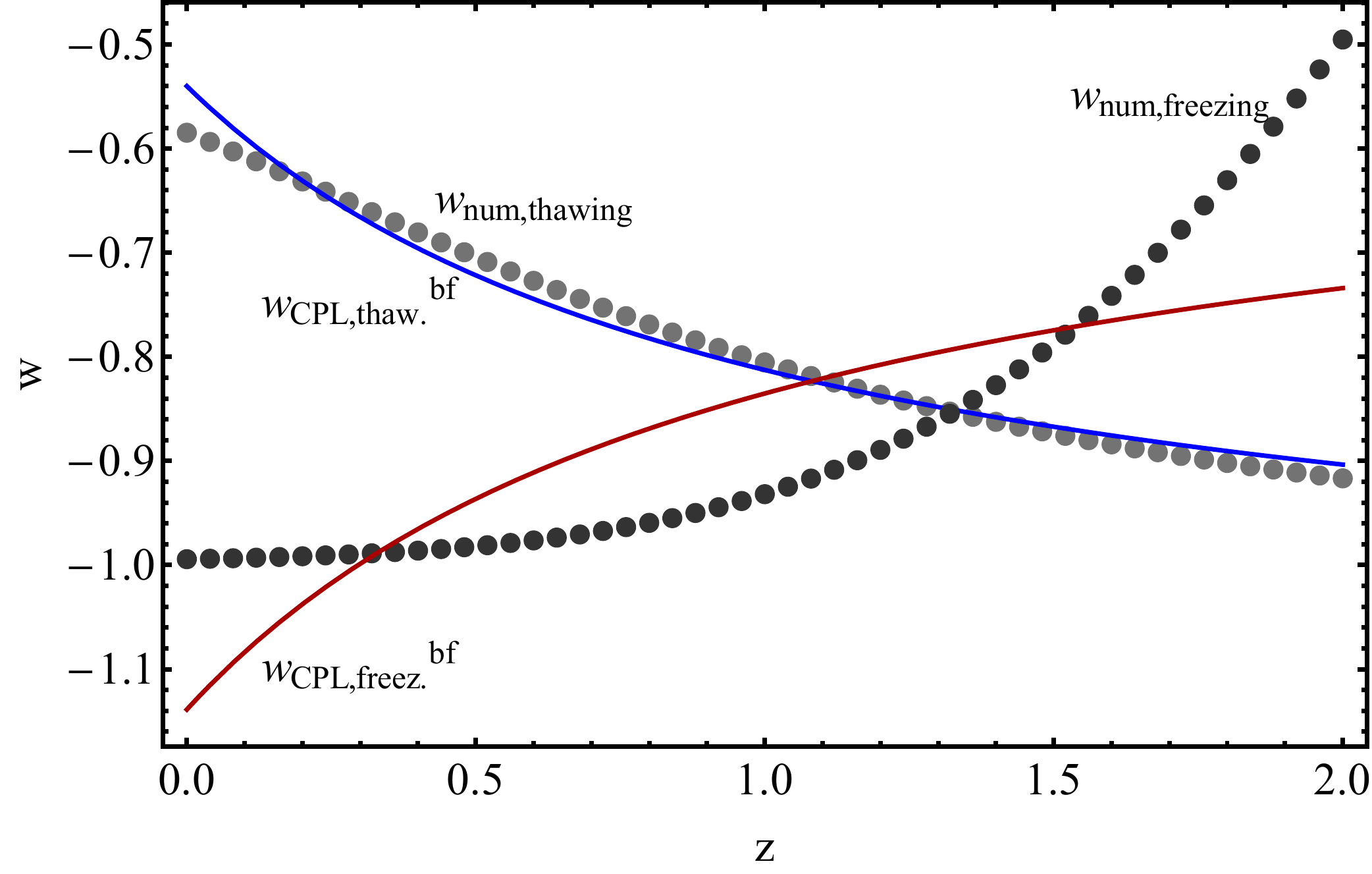}}}
\caption{Thawing and freezing quintessence models (dotted curves) with corresponding best CPL curves (continuous curves). Notice that CPL can fit thawing models much more efficiently than freezing models.}
\label{fig:Figure_1}
\end{figure}

It is easy to see that there is no two parameter parametrization that can capture the features of both types of behavior. For example the CPL parametrization is convex when decreasing and therefore in can capture the features of thawing models. However, when increasing it is concave and therefore it has difficulty to fit freezing models. This is demonstrated in Fig.~ \ref{fig:Figure_1} where we show typical forms of $w(z)$ corresponding to thawing and freezing models along with the corresponding best fit CPL parametrizations. Different types of difficulties of the CPL parametrization have also been pointed out recently in \cite{Scherrer:2015tra}.

Several parametrizations have been proposed in the literature \cite{Jassal:2004ej, Feng:2012gf, Wetterich:2004pv, Nesseris:2004wj, Linder:2006sv, Cooray:1999da, Ma:2011nc, Mukherjee:2016trt, delaMacorra:2015aqf, Barai:2004an, Choudhury:2003tj, Gong:2005de, Lee:2005id, Upadhye:2004hh, Lazkoz:2010gz, Sello:2013oja, Feng:2011zzo, Astier:2000as, Weller:2000pf, Efstathiou:1999tm, Barboza:2011gd, Cooray:1999da, Tarrant:2013xka}. As examples we refer to the Barboza-Alcaniz (BA) \cite{Barboza:2011gd} defined as \be w(z)=w_0+w_a \frac{z (1+z)}{1+z^2}, \label{ba} \ee the Jassal-Bagna-Padmanabhan \cite{Jassal:2004ej} (JBP) defined as \be w(z)=w_0+w_a \frac{z}{\left(1+z\right)^2}, \label{jbp} \ee and the linear in redshift \cite{Cooray:1999da} defined as \be w(z)=w_0+w_a z. \label{lin} \ee

\begin{figure}[!t]
\centering
\vspace{0cm}\rotatebox{0}{\vspace{0cm}\hspace{0cm}\resizebox{0.49\textwidth}{!}{\includegraphics{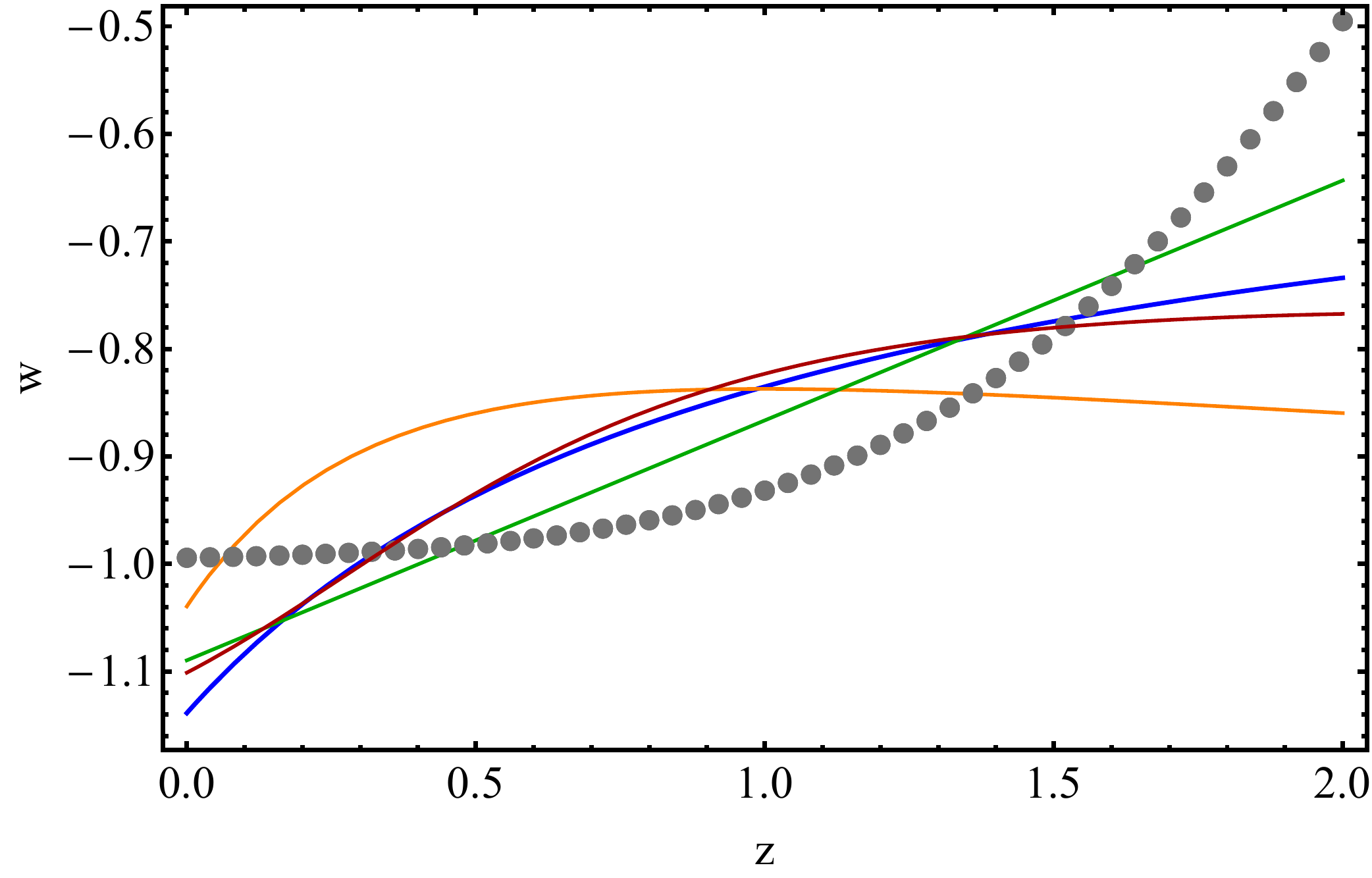}}}
\caption{Increasing forms of the CPL parametrization \eqref{cpl} (blue curve), BA \eqref{ba} (dark red curve), JBP \eqref{jbp} (orange curve) and Linear \eqref{lin} (green curve) superposed with a typical freezing $w(z)$. All these parametrizations are concave when increasing.}
\label{fig:Figure_2}
\end{figure}

Interestingly, none of these parametrizations is appropriate for fitting freezing models since they are all concave when increasing as shown in Fig.~\ref{fig:Figure_2}.

In view of the existence of two classes of behaviors for $w(z)$ (thawing and freezing) and the fact that the currently used parametrizations appear to be suitable only for the thawing class, the following questions arise
\begin{itemize}
\item[1.] What two parameter parametrization is optimal for fitting each class of models and in particular the freezing class?
\item[2.] What kind of misleading conclusions may arise when  using an inappropriate  parametrization to fit deviations from \lcdm ? Inappropriate parametrization in this context is one that does not have the concavity features corresponding to the underlying cosmological model.
\end{itemize}

The goal of the present analysis is to address these questions. The structure of this paper is the following: In Section \ref{sec:Section 2} we review thawing and freezing quintessence models and derive $w(z)$ in the context of a general but simple quintessence potential for freezing and thawing initial conditions. We also introduce a simple measure that quantifies the suitability of a parametrization for fitting a given quintessence model. We implement this measure and derive the suitability of a few parametrizations for fitting freezing and thawing forms of $w(z)$. Among the tested parametrizations is a generalized form of CPL (called nCPL) which is equipped with one more parameter ($n$) allowing it to interpolate between a two parameter freezing and a thawing parametrization.

For $n=1$ it reduces to the usual form of CPL (suitable for thawing models) while for larger values of $n$ it develops freezing properties (convex when increasing). In Section \ref{sec:Section 3} we review current observational constraints on thawing and freezing models pointing out that thawing models are significantly more constrained observationally due to their predicted divergence of $w(z)$ from the \lcdm \ value $w=-1$ at late times when there are several datasets constraining $w$ to be close to the \lcdm \ value $-1$. We also generate mock $w(z)$ scattered data with errorbars consistent with current observational constraints but with mean values obtained from the specific thawing and freezing quintessence models of Section \ref{sec:Section 2}. We fit the data obtained from observationally allowed freezing  models with  freezing  parametrizations and with thawing parametrizations (\eg CPL) showing that several incorrect statistically significant conclusions can be obtained if the inappropriate (thawing) parametrization is used. This analysis is repeated  with data obtained from observationally allowed thawing models even though this class of models has small allowed deviations from \lcdm. In Section \ref{sec:Section 4} we generate scattered luminosity distance $d_L(z)$ data consistent with {\it{Union 2.1}} errorbars but with mean values obtained from the specific thawing and freezing quintessence models of Section \ref{sec:Section 2}. We fit the data obtained from observationally allowed freezing  models with  freezing  parametrizations and with thawing parametrizations (\eg CPL) at the luminosity distance level testing the conclusions of Section \ref{sec:Section 3} with data at the luminosity distance level.

Finally in Section \ref{sec:Conclusion} we conclude and summarize  stressing  the existence of two classes of parametrizations and the importance of identifying and using  the correct class for a given dataset. We also discuss possible extensions of the present analysis.

\section{Thawing and Freezing quintessence models and corresponding parametrizations}
\label{sec:Section 2}

We assume a minimally coupled scalar quintessence field $\phi$. The action is given by \be S_\phi= \int d^4 x \sqrt{-g} \left(-\frac{1}{2}g^{\mu\nu}\D_{\mu}\phi\D_{\nu}\phi-V(\phi)\right), \ee where $V(\phi)$ is the scalar field's potential. The Lagrangian density is \be \mathcal{L}=\frac{1}{2}\dot{\phi}^2-V(\phi). \ee We may extract the field's pressure and density from the energy-momentum tensor of the action, which are \be p_{\phi}=\frac{1}{2}\dot{\phi}^2-V(\phi), \ \ \ \ \rho_{\phi}= \frac{1}{2}\dot{\phi}^2+V(\phi) \ee respectively. The scalar field obeys the dynamical equation \be \ddot{\phi}+3H\dot{\phi}+\frac{dV}{d\phi}=0, \label{KG-equation} \ee where $H$ is the Hubble parameter.

The Friedman equation in the presence of a matter fluid with equation of state $w_m$ take the form \ba H^2= \frac{\kappa^2}{3} \left(\frac{1}{2}\dot{\phi}^2+V(\phi)+\rho_m\right), \label{friedman_equation_1} \\ \dot{H}=-\frac{\kappa^2}{2} \left(\dot{\phi}^2+(1+w_m)\rho_m\right), \label{friedman_equation_2} \ea
where $\kappa=8\pi G$ is set to $1$ in natural units and $w_m=0$ for matter. In order to construct the dynamical system for the evolution of the field and the metric we define the following dimensionless quantities: \be x \equiv \frac{\dot{\phi}}{\sqrt{6}H}, \ \ y \equiv \frac{\sqrt{V}}{\sqrt{3}H}, \ \ \lambda \equiv -\frac{V_\phi}{V}, \ \ \Gamma \equiv \frac{V V_{\phi\phi}}{V_\phi^2}, \label{dimensionless_quantities} \ee
where $V_\phi$ is the first and $V_{\phi\phi}$ is the second derivative of potential with respect to the scalar field. Then equations \eqref{KG-equation}, \eqref{friedman_equation_1} and \eqref{friedman_equation_2} for the case of dark matter (\ie $w_m=0$) can be written in an autonomous form, making the following simplified autonomous system
\begin{align}
\frac{dx}{dN} &= -3x+\sqrt{\frac{3}{2}} \lambda y^2+\frac{3}{2}x \left(1+x^2-y^2\right), \label{aut_syst_1} \\
\frac{dy}{dN} &= -\sqrt{\frac{3}{2}} \lambda xy+\frac{3}{2}y \left(1+x^2-y^2\right), \label{aut_syst_2} \\
\frac{d\lambda}{dN} &= -\sqrt{6}\lambda^2 (\Gamma-1)x, \label{aut_syst_3}
\end{align}
where $N=N_0+ln(a)$ is the number of e-foldings and $N_0$ the present time. Using equations \eqref{dimensionless_quantities} we write the scalar field's density parameter as \be \Omega_\phi=x^2+y^2 \label{density_parameter} \ee and the effective equation of state of the scalar field as \be \gamma_\phi\equiv 1+w_\phi=\frac{2x^2}{x^2+y^2}. \label{effective_eos} \ee

\begin{figure}[t!]
\centering
\vspace{0cm}\rotatebox{0}{\vspace{0cm}\hspace{0cm}\resizebox{0.49\textwidth}{!}{\includegraphics{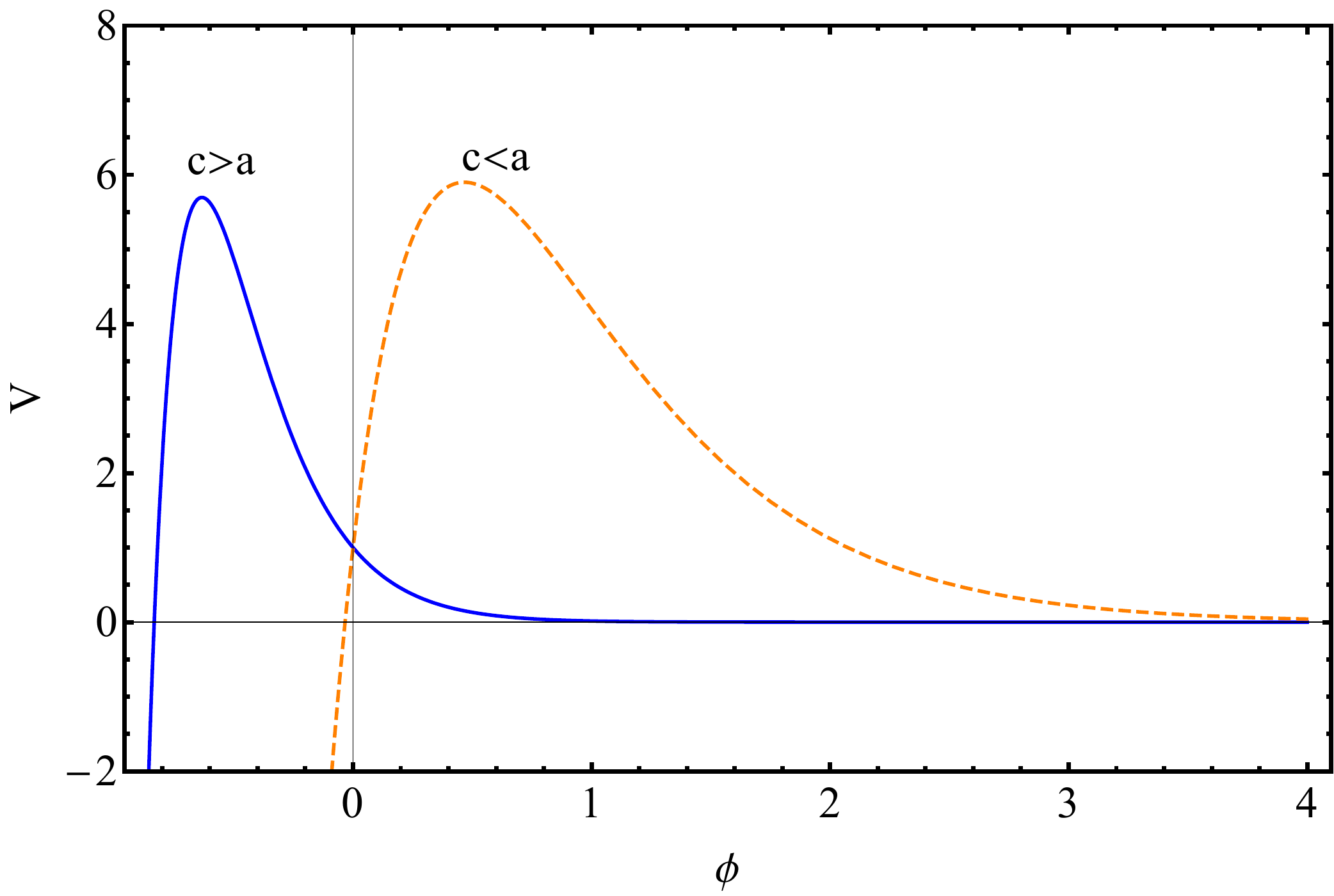}}}
\caption{The assumed scalar field potential \eqref{potential} for parameter values $c>\alpha$ (continuous line) and $c<\alpha$ (dashed line).}
\label{fig:Figure_3}
\end{figure}

We now assume a quintessence potential of the form \cite{Clemson:2008ua} \be V(\phi)=V_i \ e^{-c\phi}\left[1+\alpha\phi\right], \label{potential} \ee where c and $\alpha$ are parameters. Without loss of generality we assume an initial condition $\phi_i = 0$. Given this initial condition, it is easy to see that acceptable quintessence behavior with positive definite potential energy during evolution is obtained if $c>\alpha$. This is shown in Fig.~ \ref{fig:Figure_3} where we show a plot of the potential for $c>\alpha$ and $c<\alpha$. Notice that for $c<\alpha$ the field evolution starting from $\phi_i=0$ is unbounded from below and leads to a Big Crunch without going through an accelerating phase. Thus, in what follows we assume $c\geq\alpha$. For $c=\alpha$ the field remains constant in unstable equilibrium and the cosmological evolution is almost  identical to \lcdm \  (hilltop quintessence \cite{Dutta:2008qn, Dutta:2009dr}).

We also assume the initial condition $\dot{\phi}_i=0$ since this leads to thawing quintessence. Assuming a proper negative time derivative for the scalar field leads to a freezing quintessence behavior. With the thawing initial conditions ${\phi}_i=0$, ${\dot \phi}_i=0$, the initial condition for the dynamical variable $\lambda$ becomes \be \lambda_i=-V^\prime/V=c-a \label{eq:lambda_i} \ee and the corresponding dynamical equation for the potential \eqref{aut_syst_3} takes the form \be \frac{d\lambda}{dN}= \sqrt{6}\left(\lambda-c \right)^2 x \ee

\begin{figure}[!t]
\centering
\vspace{0cm}\rotatebox{0}{\vspace{0cm}\hspace{0cm}\resizebox{0.49\textwidth}{!}{\includegraphics{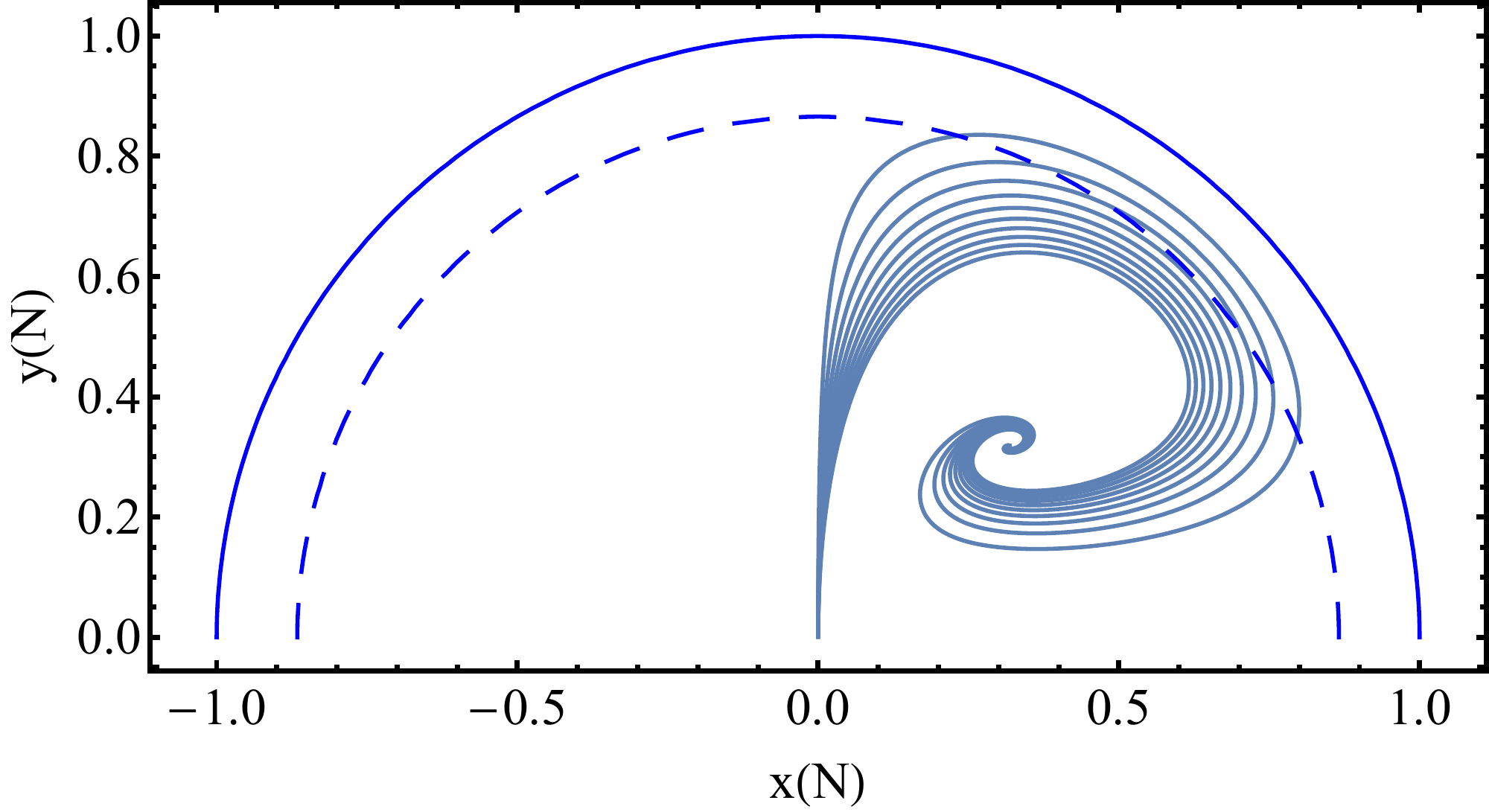}}}
\caption{Evolution for $\lambda_i= 0.1$ (\textit{outer curve}) to $1$ (\textit{inner curve}) in steps of $0.1$ for $c = 4$, with initial conditions $x_0=10^{-5}, \ y_0=10^{-3} \ $ in order to lead to thawing quintessence (${\phi}_i=0$, ${\dot \phi}_i=0$). The \textit{dashed semicircle} shows the present-day dark energy density parameter $\Omega_{\phi,0} \approx 0.75$. Obviously, the semicircle reaches the x axis at $x=\pm \sqrt{\Omega_{\phi,0}}\simeq0.866$. Trajectories that reach this may agree to the evolution of our universe. The origin is the initial point at some early time during matter domination, and the \textit{unit semicircle} shows scalar field domination ($\Omega_\phi=1$).}
\label{fig:Figure_4}
\end{figure}

We now solve the dynamical system with thawing initial conditions defining the present time (corresponding to $N=N_0$) as the time when $\Omega_\phi = 0.75$ for the first time. Note that it is not possible to get to the present value of $\Omega_\phi$ for all initial conditions. This is demonstrated in Fig.~\ref{fig:Figure_4} where we show the solution of the dynamical system for various initial conditions.

The equation of state parameter $w(N)$ can be determined from the solution $x(N)$, $y(N)$ as
\be w(N)=\frac{x^2(N) - y^2(N)}{x^2(N)+y^2(N)}, \label{wnum}
\ee while the corresponding scale factor (rescaled so that $a(N_0)=1$) is of the form \be a=\frac{1}{1+z}=e^{N-N_0}. \label{scale_factor} \ee

The numerically obtained form of the equation of state parameter  $w(N)$ may now be converted to a function of redshift $w(z)$ using equations \eqref{wnum}, \eqref{scale_factor} and fit using specific parametrizations for $w(z)$ like the CPL parametrization. In the context of such a fit, the parameters $w_0$ and $w_a$ of equation \eqref{cpl} are determined by a least squares fit to the numerically obtained $w(z)$. A measure of the quality of fit may be defined in analogy to the likelihood for random data as discussed below.

Previous studies \cite{Clemson:2008ua, Marsh:2014xoa} have used an expansion of the numerically obtained $w(a)$ (obtained from equations \eqref{wnum} and \eqref{scale_factor}) around $a=1$ to find the predicted parameter values $w_0$ and $w_a$. This approach does not involve a fit over a finite range of redshifts ($z$ or of scale factor $a$) and therefore the derived form of the parametrized $w_\text{cpl}(z)$ agrees with the numerically obtained $w(z)$ only in the limit of low $z$) (or $a=1$).

The approach of these studied involving expanding around the present time, using equation \eqref{scale_factor} and the CPL parametrization \be w(N)=w_0+w_a (1-e^{N-N_0}). \label{cpl(N)} \ee

Thus, we find the CPL parameters
\be w_0^\text{lin.} = \frac{x_0-y_0}{x_0+y_0} \label{w_0} \ee and
\be \begin{split} w_a^\text{lin.} &= -\left.\frac{dw}{da}\right\rvert_{a \rightarrow 1}= -\left(\frac{dw}{dN}e^{-N+N_0}\right)_{N \rightarrow N_0} \ra \\ w_a^\text{lin.} &=\frac{4x_0 y_0}{\left(x^2_0+y^2_0\right)^2} \left(3x_0 y_0- \lambda_0 \sqrt{\frac{3}{2}}y^3_0 - \lambda_0 \sqrt{\frac{3}{2}} x^2_0 y_0\right), \label{w_a} \end{split} \ee
where $x_0$, $y_0$ and $\lambda_0$ correspond to present day values ($N$=$N_0$). These values of parameters provide good fit to the numerically obtained $w(z)$ only in the limit of $z \rightarrow 0$.

\begin{figure}[!t]
\centering
\vspace{0cm}\rotatebox{0}{\vspace{0cm}\hspace{0cm}\resizebox{0.49\textwidth}{!}{\includegraphics{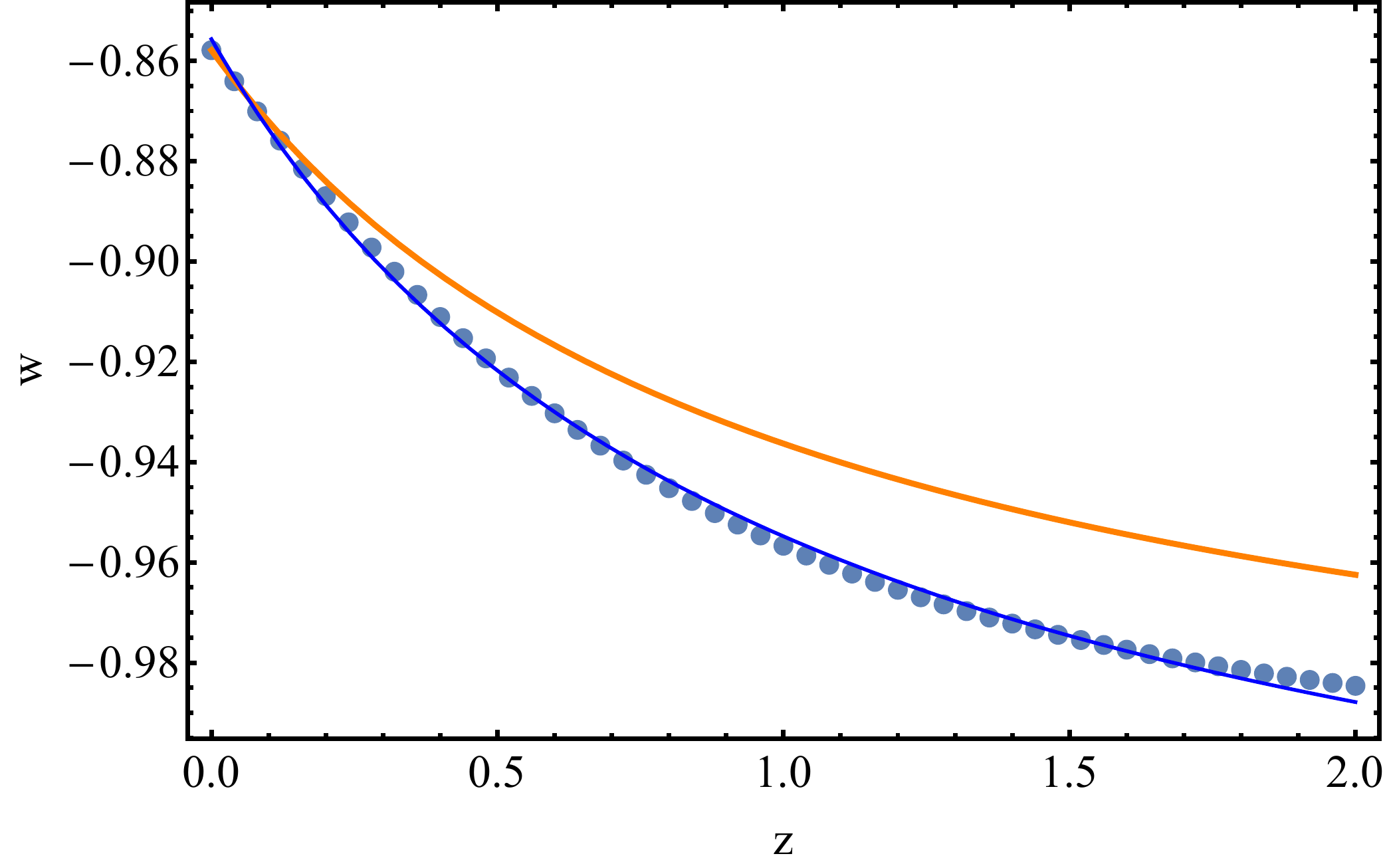}}}
\caption{The numerical values of $w$ (points), the CPL linear fit (orange thick line) and the CPL best fit of $w(z)$ (blue line), obtained for $c=1, a=0.1$.}
\label{fig:Figure_5}
\end{figure}

To demonstrate this fact, we show in Fig.~\ref{fig:Figure_5} a plot of the numerically obtained $w(z)$ (blue dots) along with the $w_\text{cpl}^\text{lin}(z)\equiv w_0^\text{lin.}+w_a^\text{lin.} \frac{z}{1+z}$ (orange dashed line) obtained from the linear expansion around the present time (equations \eqref{w_0} and \eqref{w_a}) and the $w_\text{cpl}^\text{bf}(z)$  obtained from a least squares best fit of the CPL parametrization (equation \eqref{cpl}) to the numerical result in the redshift range $z\in [0,2]$. Clearly, the parameter values obtained with the least squares fit are a much better representation of the quintessence model than the parameter values obtained using the expansion up to linear order. As expected all three forms of $w(z)$ agree in the limit of very low redshift ($z\ll 1$).

In order to quantify the range of redshifts where the linear approximation $w_\text{cpl}^\text{lin}(z)$ used in previous analyses \cite{Clemson:2008ua, Dutta:2009dr} is in agreement with the more accurate best fit $w_\text{cpl}^\text{bf}(z)$ we define the parameters $w_0^\text{bf}(z_\text{max})$ and $w_a^\text{bf}(z_\text{max})$ as the best fit parameter values corresponding to $w_\text{cpl}^\text{bf}(z)$ fitted in the redshift range $z\in [0,z_\text{max}]$. The deviation of these parameters from the corresponding ones obtained in the linear approximation of the numerical results (equations \eqref{w_0} and \eqref{w_a}) is quantified through the quantities
\begin{align}
w_{0,\text{dev}}(z_\text{max}) &= \frac{2\left(w_0^\text{bf}(z_\text{max})-w_0^\text{lin}\right)}{w_0^\text{lin}+w_0^\text{bf}(z_\text{max})}, \label{w0dev} \\
w_{a,\text{dev}}(z_\text{max}) &= \frac{2\left(w_a^\text{bf}(z_\text{max})-w_a^\text{lin} \right)}{w_a^\text{lin}+w_a^\text{bf}(z_\text{max})}. \label{wadev}
\end{align}

\begin{figure*}[!t]
\centering
\vspace{0cm}\rotatebox{0}{\vspace{0cm}\hspace{0cm}\resizebox{0.99\textwidth}{!}{\includegraphics{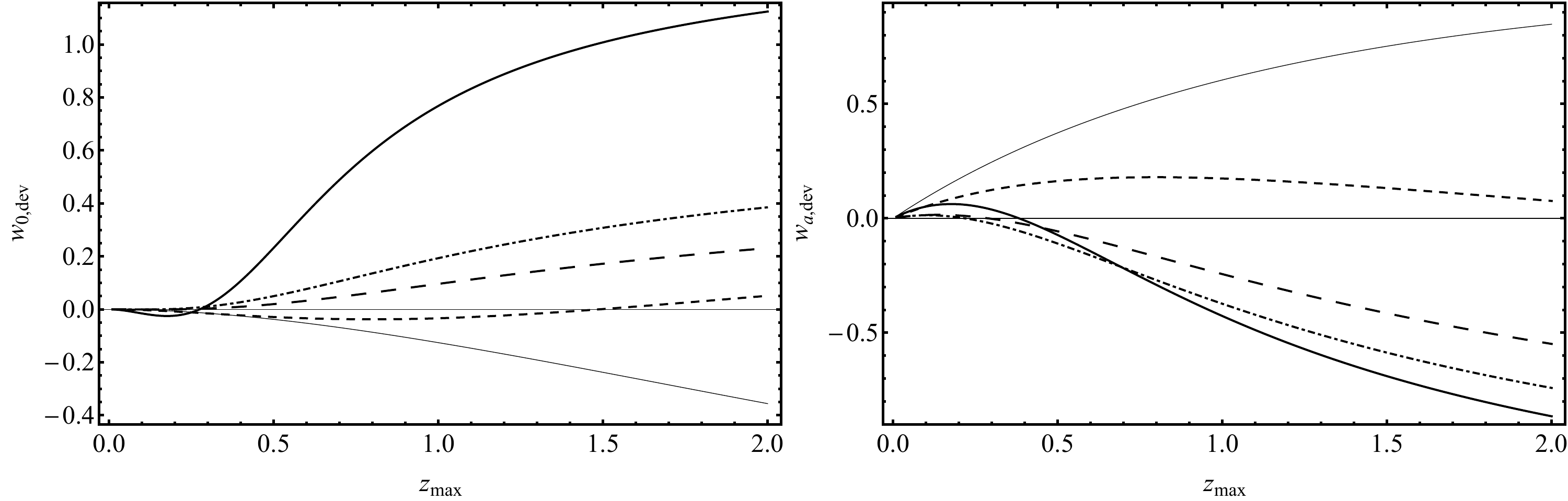}}}
\caption{The form of $w_{0,dev}(z)$ and $w_{a,dev}(z)$ for the CPL parametrization for different values of $c,\alpha$, \ie for $c=2, a=0.1$ (thick line), $c=3, a=2$ (short dashed line), $c=3.5, a=3$ (long dashed line), $c=4, a=3.7$ (dot-dashed line) and $c=5, a=4.9$ (solid line).}
\label{fig:Figure_6}
\end{figure*}

In Fig.~\ref{fig:Figure_6} we show the dependence of $w_{0,\text{dev}}$, $w_{a,\text{dev}}$ on the maximum redshift $z_\text{max}$ used for the fit, for various values of parameters of the quintessence potential. Notice that for $z_\text{max}<0.2$ the deviation is less than a few percent. However, for larger $z_\text{max}$ the deviation can be as large as $100\%$. It is therefore important to avoid using the linear approximation when estimating the CPL parameter values that describe a quintessence model.

\begin{figure}[!t]
\centering
\vspace{0cm}\rotatebox{0}{\vspace{0cm}\hspace{0cm}\resizebox{0.49\textwidth}{!}{\includegraphics{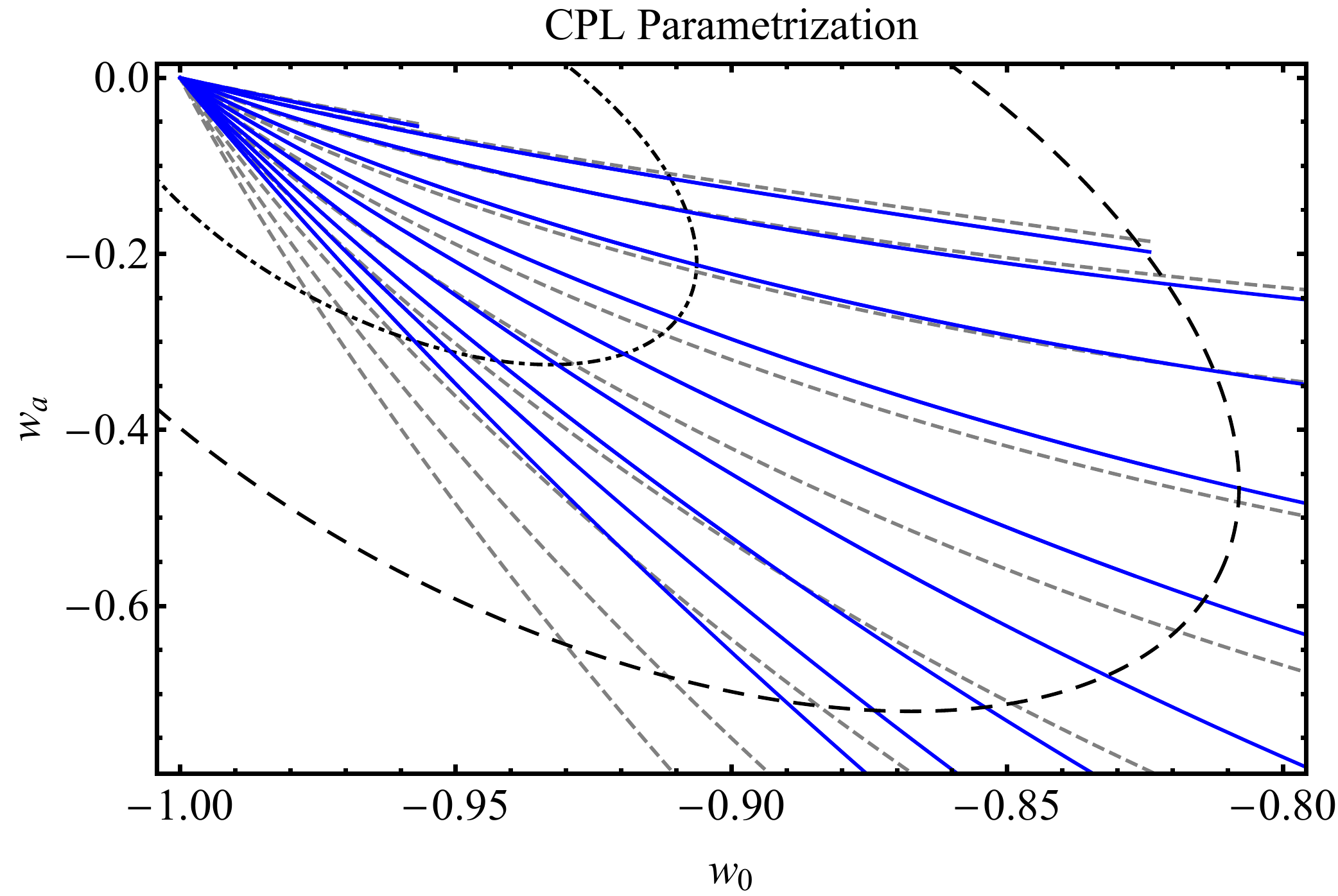}}}
\caption{CPL linear $w_0^\text{lin.}-w_a^\text{lin.}$ plane (dashed lines) from $c=0.5$ (top dashed line) to $c=5.0$ (lower dashed line) in steps of 0.5 and similarly CPL best fit $w_0^\text{bf.}-w_a^\text{bf.}$ plane (solid lines). The dot-dashed line indicates the $1\sigma \ (\sim 68\%)$, while the long-dashed line indicated the $2\sigma \ (\sim 95\%)$ of observational constraints \cite{Komatsu:2008hk}.}
\label{fig:Figure_7}
\end{figure}

The estimated CPL parameter values using both the linear (dashed lines) and the best fit (solid lines) approach are shown in Fig.~\ref{fig:Figure_7} for a wide range of parameters $c$, $\alpha$ of the quintessence potential. In the same plot we also show an estimate of observational constraints on the CPL parameters at the $1\sigma$ and the $2\sigma$ level \cite{Komatsu:2008hk}. Notice the significant deviation between the solid and the dashed lines indicating the significant difference between linear and the best fit approaches in estimating the CPL parameter values corresponding to a given quintessence model.

We now use the best fit approach to derive the parameter values of a parametrization fitting the numerically obtained $w(z)$ and compare the quality of fit provided by different parametrizations to thawing and freezing models. We define a simple measure $q$ of the quality of fit as \be q=\int_0^{z_\text{max}} \frac{\left|w_\text{num}(z)-w_\text{par.}^\text{bf}(z)\right|}{z_\text{max}} \ dz. \label{quality} \ee
For a perfect fit $q=0$ and $q$ increases as the quality of fit decreases.

We use the thawing initial conditions (${\dot \phi}_i=0$) and implement the measure $q$ to compare the quality of fit of various parametrizations to thawing quintessence $w(z)$. We compare the quality of fit of the following parametrizations
\begin{itemize}
\item[i.] CPL ($w_\text{CPL}(z)$ \eqref{cpl})
\item[ii.] BA ($w_\text{BA}(z)$ \eqref{ba})
\item[iii.] Linear ($w_\text{Linear}(z)$ \eqref{lin})
\item[iv.] Sqrt, proposed in the present study and defined as \be w_\text{Sqrt}(z)=w_0+w_a \frac{z}{\sqrt{1+z^2}} \label{sqrt} \ee
\item[v.] Generalized CPL (nCPL) defined as
\be  w_\text{nCPL}(z)=w_0+w_a (1-a)^n=w_0 + w_a \left(\frac{z}{1+z}\right)^n \label{ncpl} \ee
\end{itemize}

\begin{figure*}
\centering
\vspace{0cm}\rotatebox{0}{\vspace{0cm}\hspace{0cm}\resizebox{0.99\textwidth}{!}{\includegraphics{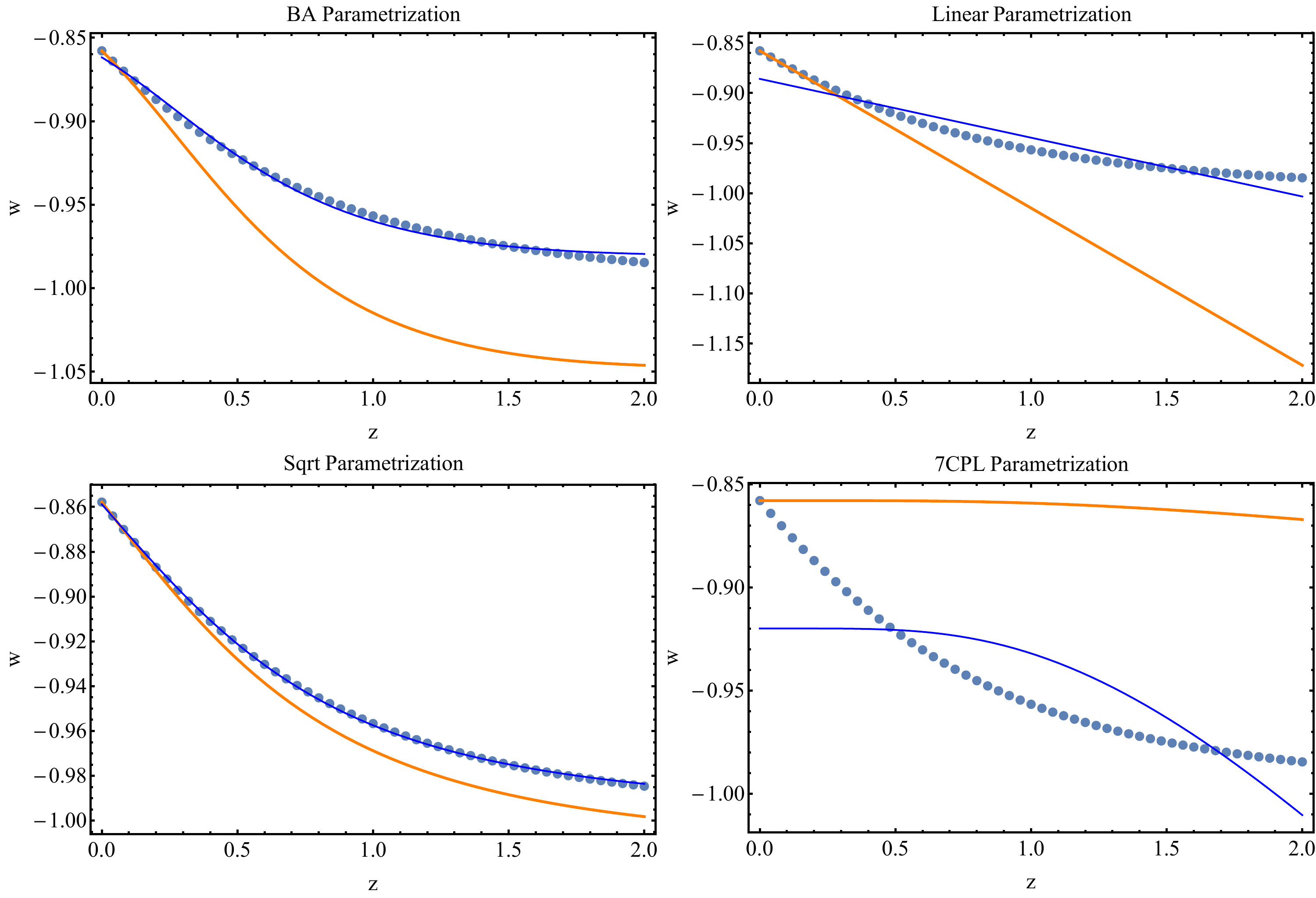}}}
\caption{The numerical values of $w(z)$ (points) in thawing model  for $c=1$ and $a=0.1$ superposed on each parametrization's linear fit (orange thick line) and each parametrization's best fit of $w(z)$ (blue line).}
\label{fig:Figure_8}
\end{figure*}

The nCPL parametrization interpolates between being concave to being convex when increasing as the value of $n$ from $n=1$ (CPL) to larger $n$. Therefore as discussed in the Introduction \ref{sec:Introduction} for larger $n$ it is more suitable for fitting freezing models while for $n=1$ it reduces to the usual CPL parametrization which is more suitable for fitting thawing models. This is demonstrated in Fig.~\ref{fig:Figure_8} where we show the numerically obtained $w(z)$ for thawing initial conditions (blue dots) along with best fit forms $w^\text{bf}(z)$ of the above parametrizations (except CPL which is shown in Fig.~\ref{fig:Figure_5}) and the corresponding forms obtained with the linear expansion $w^\text{lin}(z)$ as described above.

Clearly, parametrizations {\it{i-iv}} can provide a good fit to the thawing behavior of the numerically obtained $w(z)$. On the other hand the parametrization {\it{v}} (for $n=7$) is unable to provide a good fit to the thawing quintessence model. As it will be discussed below it is more suitable to represent freezing quintessence and other similar models with convex (when increasing) models.

Using the measure $q$ defined in equation \eqref{quality} we may quantify the quality of fit of the above parametrizations for thawing quintessence models. We thus evaluate the mean value of $q$ for each one of the above parametrizations over several parameters $c$, $\alpha$ of the quintessence potential using $z_\text{max}=2$. We consider the parameter range that corresponds to the region between the $1\sigma$ and the $2\sigma$ contours of Fig.~\ref{fig:Figure_7} so that they are consistent with observational constraints but at the same they are not very close to \lcdm \ where all parametrizations perform equally well since they include \lcdm \ as a special case.

\begin{figure}[!t]
\centering
\vspace{0cm}\rotatebox{0}{\vspace{0cm}\hspace{0cm}\resizebox{0.49\textwidth}{!}{\includegraphics{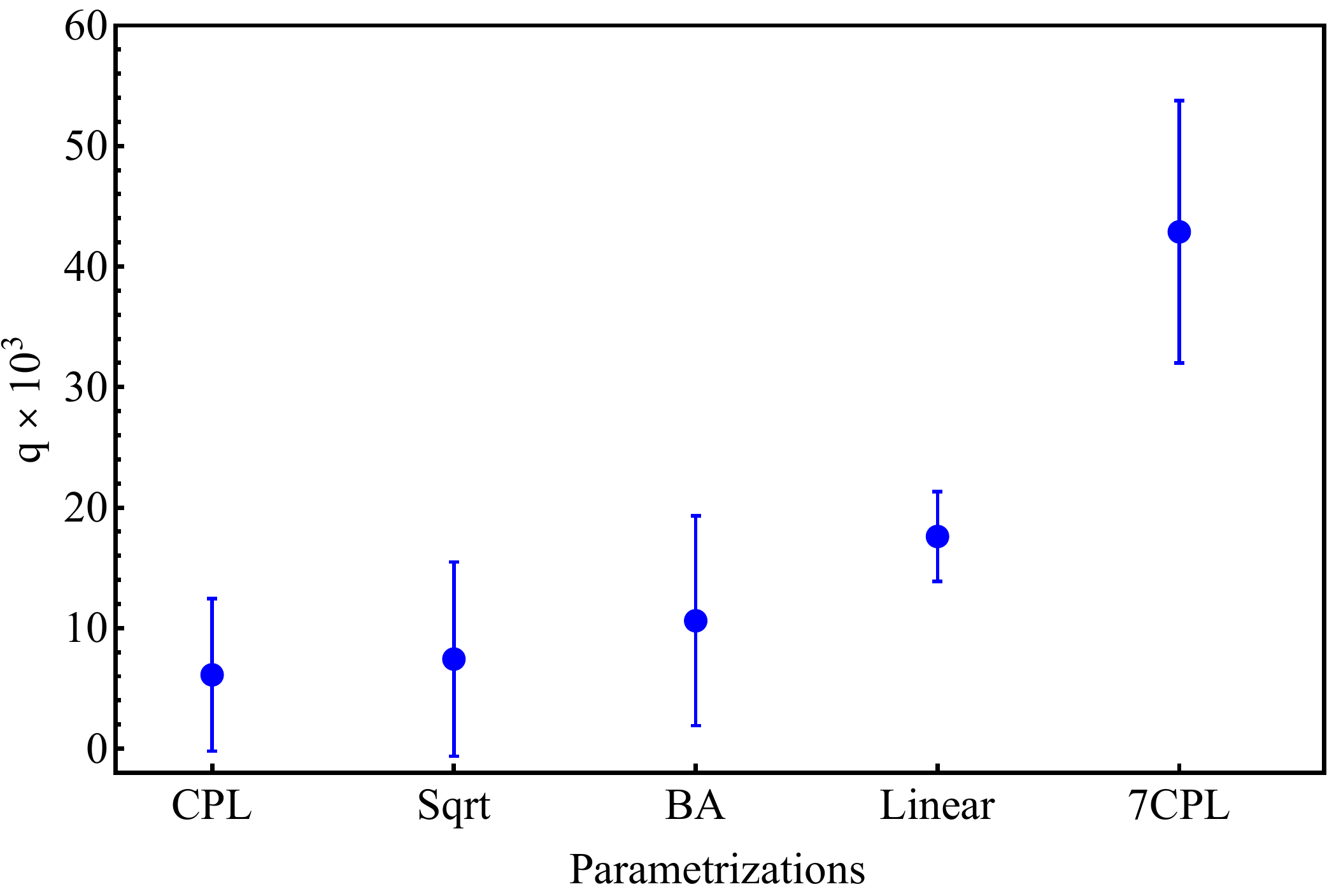}}}
\caption{Each parametrization's measure $q$ \eqref{quality} range in the underlying thawing model with initial conditions $x_i=10^{-5}, y_i=10^{-3}$, for $c$ in the range 1 to 2 and $\alpha$ in the range 0 to $c-1$ in order to get the range $\left[1\sigma,2\sigma \right]$ observational constraints parameters.}
\label{fig:Figure_9}
\end{figure}

The mean values of $q$ for each parametrization along with the $1\sigma$ deviation are shown in Fig.~\ref{fig:Figure_9}. Clearly parametrizations {\it{i-iii}} are well suited for representing thawing models since they are convex when decreasing in agreement with the thawing behavior. The {\it{Linear}} parametrization {\it{iv}} is neither convex nor concave and therefore it provides a worse quality of fit for thawing models (increased $q$). Finally, the parametrization {\it{v}} (nCPL with $n=7$) is concave when decreasing and is unable to capture the thawing behavior. It has the highest value of $q$ as shown in Fig.~\ref{fig:Figure_9}.

\begin{figure*}[!t]
\centering
\vspace{0cm}\rotatebox{0}{\vspace{0cm}\hspace{0cm}\resizebox{0.99\textwidth}{!}{\includegraphics{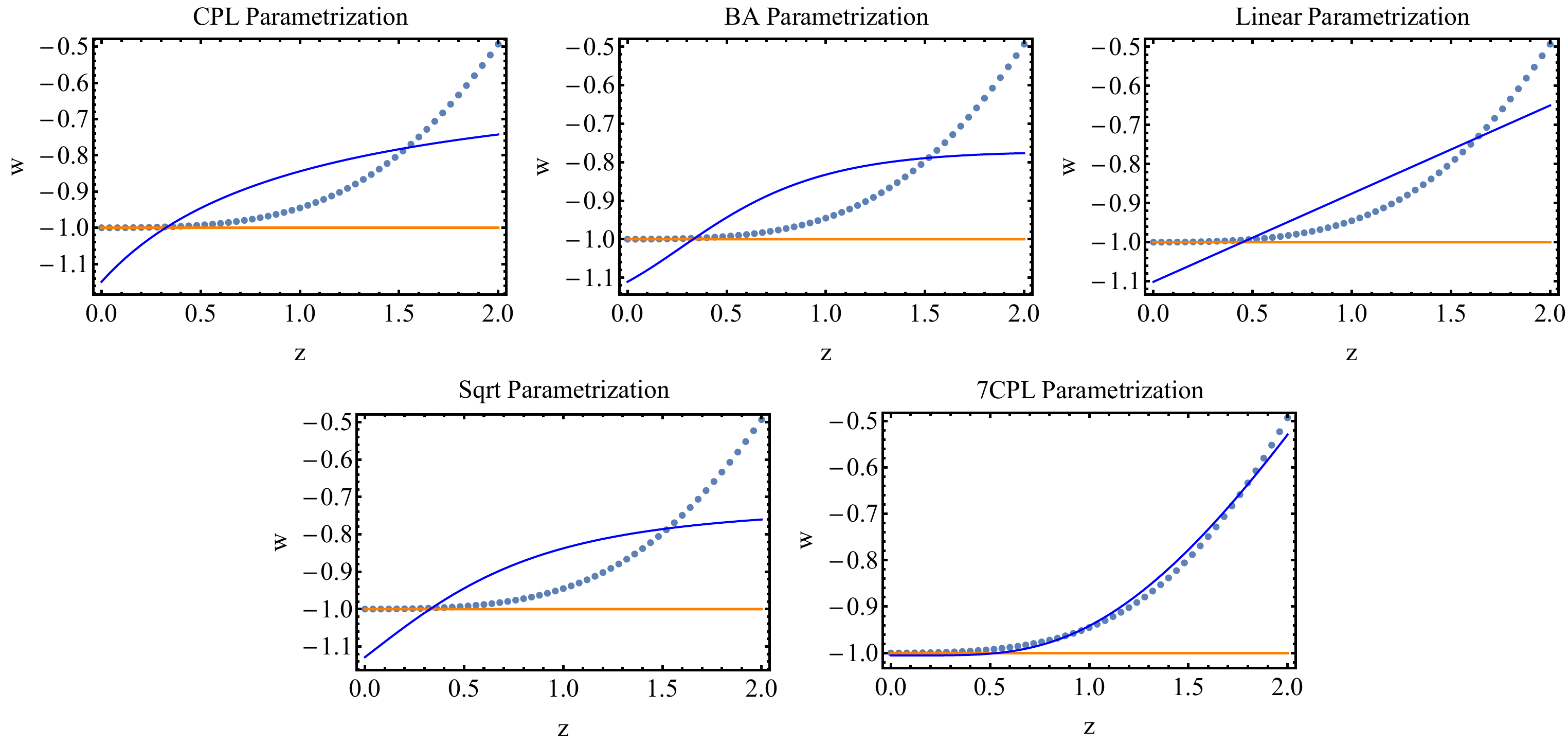}}}
\caption{The numerical values of $w$ (points) in freezing model, each parametrization's linear fit (orange thick line) and each parametrization's best fit of $w(z)$ (blue line) for $c=0.7$ and initial value $\lambda_i=-0.5$.}
\label{fig:Figure_10}
\end{figure*}

\begin{table*}[!t]
\centering
\resizebox{0.9\textwidth}{!}{
\begin{tabular}{|c|c|c|c|c|}
\hhline{=====}
Name & Parametrization & Type & $q_\text{th.}\times 10^3$ & $q_\text{fr.}\times 10^3$ \\ \hhline{=====}
Sqrt & $w(z)=w_0+w_a \frac{z}{\sqrt{1+z^2}}$ & Thawing & $0.5$ & $78.7$ \\
CPL \cite{Chevallier:2000qy, Linder:2002et} & $w(z)=w_0+w_a \frac{z}{1+z}$ & Thawing & $1.3$ & $74.9$ \\
BA \cite{Barboza:2011gd} & $w(z)=w_0+w_a \frac{z (1+z)}{1+z^2}$ & Thawing & $2.1$ & $81.7$ \\
Sine \cite{Lazkoz:2010gz} & $w(z)=w_0+w_a \sin(z)$ & Thawing & $3.6$ & $82.6$ \\
Logarithmic \cite{Feng:2011zzo} & $w(z)=w_0+w_a \ln(1+z)$ & Thawing & $5.2$ & $62.8$ \\
MZ Model 1 \cite{Ma:2011nc} & $w(z)=w_0+w_a \left(\frac{\ln(2+z)}{1+z}-\ln 2 \right)$ & Thawing & $5.5$ & $62.1$ \\
FSLL Model 2 \cite{Feng:2012gf} & $w(z)=w_0+w_a \frac{z^2}{1+z^2}$ & Thawing & $5.9$ & $64.3$ \\
Linear \cite{Cooray:1999da} & $w(z)=w_0+w_a z$ & Thawing & $9.3$ & $49.5$ \\
MZ Model 2 \cite{Ma:2011nc} & $w(z)=w_0+w_a \left(\frac{\sin(1+z)}{1+z}-\sin(1) \right)$ & Thawing & $9.7$ & $48.5$ \\
FSLL Model 1 \cite{Feng:2012gf} & $w(z)=w_0+w_a \frac{z}{1+z^2}$ & Thawing & $17.0$ & $108.5$ \\
JBP \cite{Jassal:2004ej} & $w(z)=w_0+w_a \frac{z}{\left(1+z\right)^2}$ & Thawing & $18.0$ & $108.7$ \\[0.15cm]
7CPL & $w(z)=w_0+w_a \left(\frac{z}{1+z}\right)^7$ & Freezing & $18.7$ & $8.4$ \\ \hhline{=====}
\end{tabular}}
\caption{A comparison of a variety of two parameter $w(z)$ parametrizations with respect to their ability to efficiently fit thawing and freezing quintessence models. This efficiency is measures by the $q-$value defined by equation \eqref{quality}. $q_\text{th.}$ ($q_\text{fr.}$) is the $q-$value for fitting a thawing (freezing) quintessence model.}
\label{tab:Table_1}
\end{table*}

\begin{figure}[!t]
\centering
\vspace{0cm}\rotatebox{0}{\vspace{0cm}\hspace{0cm}\resizebox{0.49\textwidth}{!}{\includegraphics{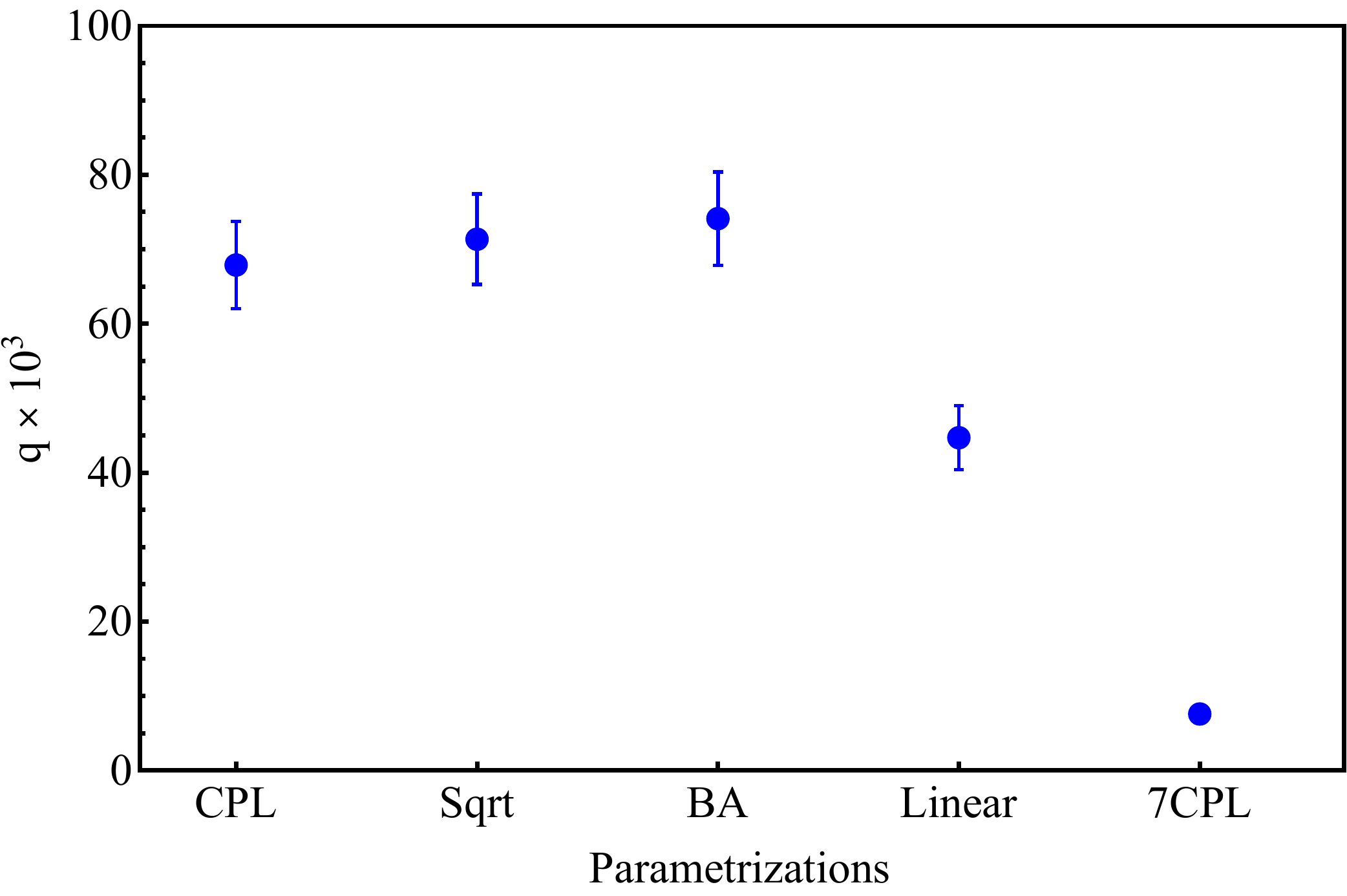}}}
\caption{Each parametrization's measure $q$ \eqref{quality} range in the underlying freezing model with $x_i=-0.5, \ y_i=0.1, \ \lambda_i=-0.5$, for $c$ in the range 0 to 0.7 in order to get the range $\left[1\sigma,2\sigma \right]$ observational constraints parameters.}
\label{fig:Figure_11}
\end{figure}

\begin{figure}[!t]
\centering
\vspace{0cm}\rotatebox{0}{\vspace{0cm}\hspace{0cm}\resizebox{0.49\textwidth}{!}{\includegraphics{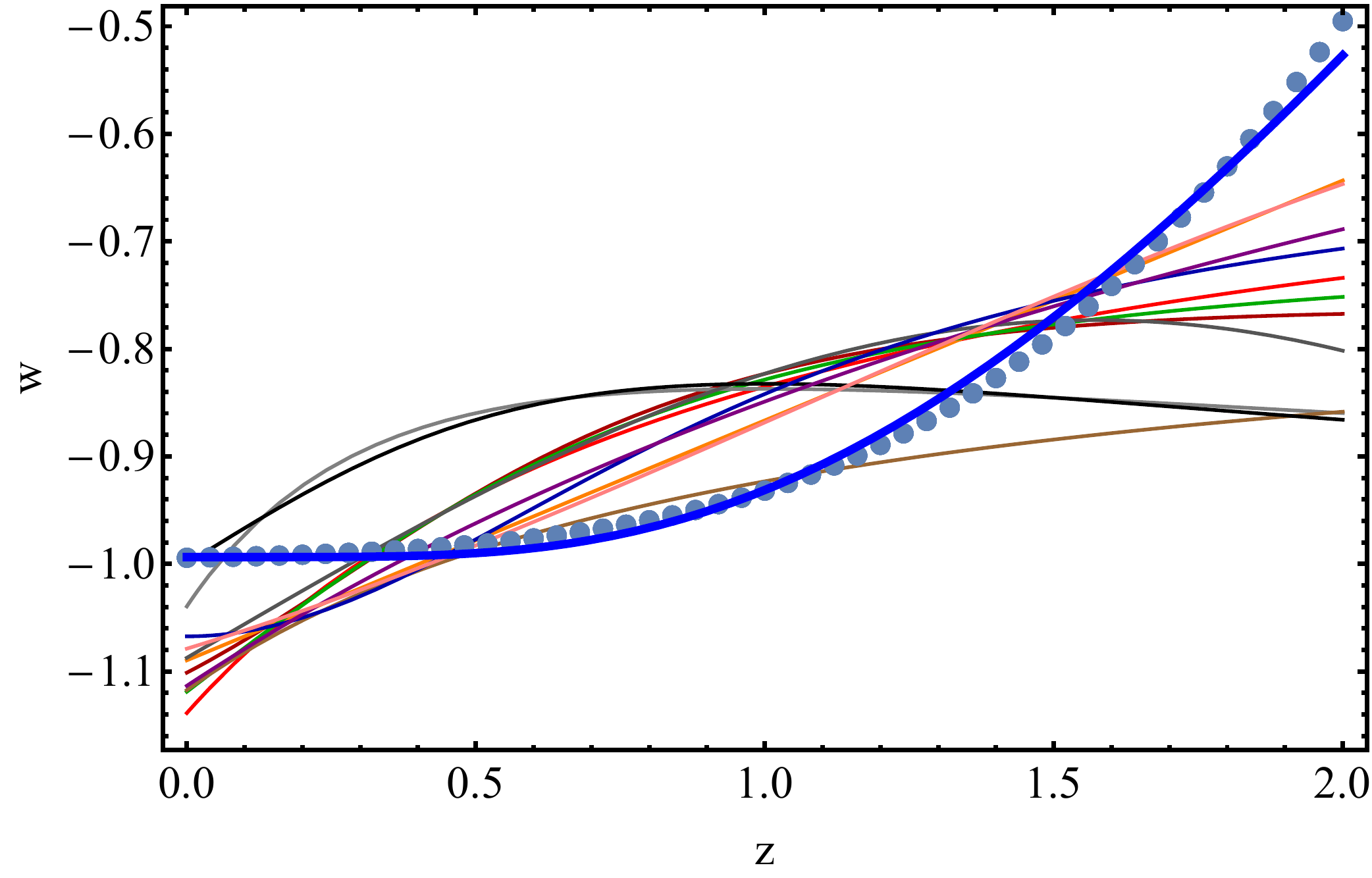}}}
\caption{The least squares best fit forms of the $w(z)$ two parameter parametrizations of Table~\ref{tab:Table_1} with respect to a typical freezing model (dotted line). Only the nCPL parametrization (blue line) with large value of $n$ ($n=7$ is used here) is able to efficiently fit an underlying freezing model.}
\label{fig:Figure_12}
\end{figure}

In order to test the ability of the parametrizations considered to fit freezing quintessence behavior we may keep the same potential of equation \eqref{potential} but change the initial conditions for the field time derivative from ${\dot \phi}_i=0$ to ${\dot \phi}_i <0$. Thus the field is initially moving up its potential while decelerating and reducing its kinetic term. The corresponding behavior for $w(z)$ is to be larger than $-1$ at early times due to the presence of the kinetic term but tend to $-1$ at late times as the kinetic term is reduced. Even though this type of behavior is not tracking at early times it does produce a freezing behavior for $w(z)$ which is analogous to the tracking-freezing form.

In Fig.~\ref{fig:Figure_10} we show the numerically obtained $w(z)$ with freezing initial conditions along with the best fit for each parametrization and the corresponding linear expansion around $z=0$ obtained with each parametrization. Notice that the linear expansion around the present time is practically indistinguishable from \lcdm \ and is not representative of the underlying freezing model.

Clearly, the thawing parametrizations {\it{i-iv}} are unable to properly represent the freezing behavior of the quintessence model. In contrast the freezing parameterization 7-CPL provides an excellent fit to the freezing model.

This is confirmed quantitatively in Fig.~\ref{fig:Figure_11} where we show the value of the measure $q$ for each parametrization, obtained for a range of potential parameter values consistent with observations,  at the $2\sigma$ level, with freezing initial conditions. Clearly, $q$ in this case behaves in the opposite manner compared to the behavior of Fig.~ \ref{fig:Figure_9}. The freezing parameterization (7CPL) has the best quality of fit (lowest $q$) while the thawing parametrizations {\it{i-iii}} are unable to represent these freezing quintessence models. The linear parametrization {\it{iv}} is unable to represent any of the two classes of models.

As it can be seen from Fig.~\ref{fig:Figure_10}, the attempt to fit the underlying freezing model with an inappropriate (thawing) parametrization like the CPL leads to misleading conclusions. First, there is a false trend for crossing of the phantom divide line $w=-1$ leading to phantom behavior which is not present in the underlying model (blue dots). In addition the gradient of $w(z)$ is significantly different from the gradient of the underlying model in both low and in high redshifts indicating a false type of future evolution for $w(z)$.

All the parametrizations we have discussed so far with the exception of nCPL (for $n>1$) belong to the thawing class and vary linearly with $z$ at low $z$. Due to this linear dependence and the small number of parameters (2), these parametrizations have difficulty to fit functional forms $w(z)$ that vary as $z^n$ with $n>1$ at low $z$ as is the case for freezing-tracking quintessence models. In fact all the two parameter parametrizations that we were able to find in the literature are suffering from a similar problem and are thus unable to efficiently fit freezing models.

In Table~\ref{tab:Table_1} we show an extensive list of such parametrizations (including the ones above) appearing in previous studies (\ie the Larkoz-Salzano-Sendra sine parametrization \cite{Lazkoz:2010gz}, the Feng-Lu logarithmic one \cite{Feng:2011zzo}, two models proposed by Ma-Zhang \cite{Ma:2011nc} and two other models proposed by Feng-Shen-Li-Li \cite{Feng:2012gf}) along with their $q-$value with respect to a typical thawing quintessence model ($q_\text{th.}$) and with respect to a typical freezing quintessence model ($q_\text{fr.}$). Almost all parametrizations (except nCPL) belong to the thawing class and are able to fit well a thawing underlying model (they have a low value of $q_\text{th.}$). However, the only parametrization with good fit to a freezing model (low $q_\text{fr.}$ value) is the nCPL parametrization with a high value of $n$ (we have used $n=7$ here but other similar values of $n$ are also providing similar good fits). This is also demonstrated in Fig.~\ref{fig:Figure_12} where we show the best fir forms of all parametrizations to a typical freezing quintessence model (dotted line) obtained by numerically solving the dynamical system \eqref{aut_syst_1}$-$\eqref{aut_syst_3} with freezing initial conditions for parameter values consistent with observations at the $2\sigma$ level.

In view of the fact that $w(z)$ parametrizations appear to be divided in two distinct classes (interpolated by the Linear parametrization {\it{iv}} of equation \eqref{lin}), the following question arises:

Given a cosmological dataset expressed as $w(z_i)$ with $1\sigma$ errors what is the optimal class of parametrizations to use in order to avoid misleading conclusions about the estimated behavior of the underlying cosmological model and uncover possible improved quality of fit with respect to \lcdm ?

The answer to this question may be obtained by implementing parametrizations from both classes and evaluating the quality of fit to the cosmological data. This procedure is demonstrated in the following section where we consider simulated mock scattered data based on underlying thawing or freezing models with errors corresponding to real data.

\section{Comparison of parametrizations using simulated datasets at the level of $w(z)$}
\label{sec:Section 3}

\begin{figure}[!t]
\centering
\vspace{0cm}\rotatebox{0}{\vspace{0cm}\hspace{0cm}\resizebox{0.49\textwidth}{!}{\includegraphics{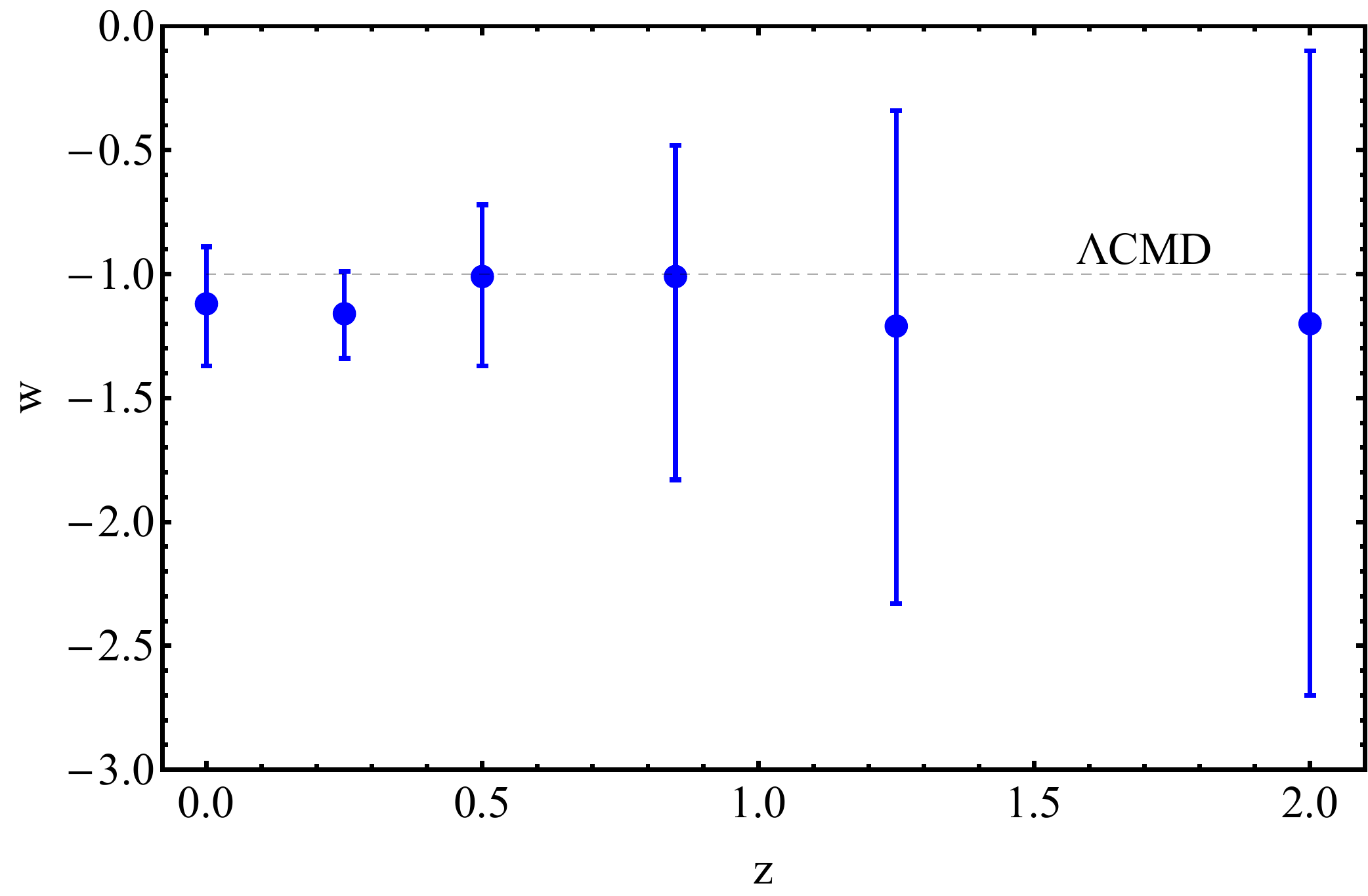}}}
\caption{Recent constraints on $w(z)$ from a WMAP+SNLS+BAO+H(z) dataset \cite{Said:2013jxa}.}
\label{fig:Figure_13}
\end{figure}

\begin{figure*}[!t]
\centering
\vspace{0cm}\rotatebox{0}{\vspace{0cm}\hspace{0cm}\resizebox{0.99\textwidth}{!}{\includegraphics{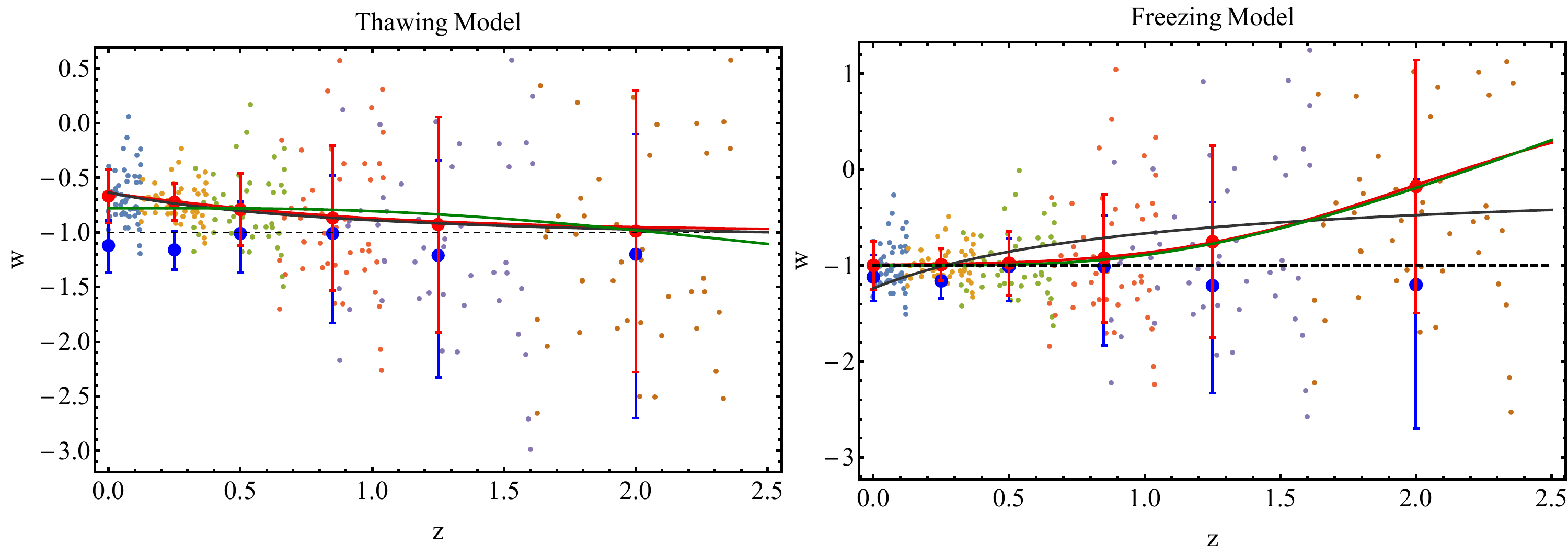}}}
\caption{Thawing (left) and freezing (right) model scattered Gaussian mock data (50 data in each bin with different color). Besides the constraints of Fig.~\ref{fig:Figure_13} (blue errorbars), we also include the underlying model (red line) along with the constraints corresponding to those of Fig.~\ref{fig:Figure_13} (red errorbars), best fit CPL parametrization (dark gray line), best fit 7CPL parametrization (green line) and the \lcdm \ line (dashed). Notice on the left, the best fit CPL is almost identical with the underlying thawing model, while on the right the best fit 7CPL is identical with the underlying freezing model.}
\label{fig:Figure_14}
\end{figure*}

\begin{figure*}[!t]
\centering
\vspace{0cm}\rotatebox{0}{\vspace{0cm}\hspace{0cm}\resizebox{0.99\textwidth}{!}{\includegraphics{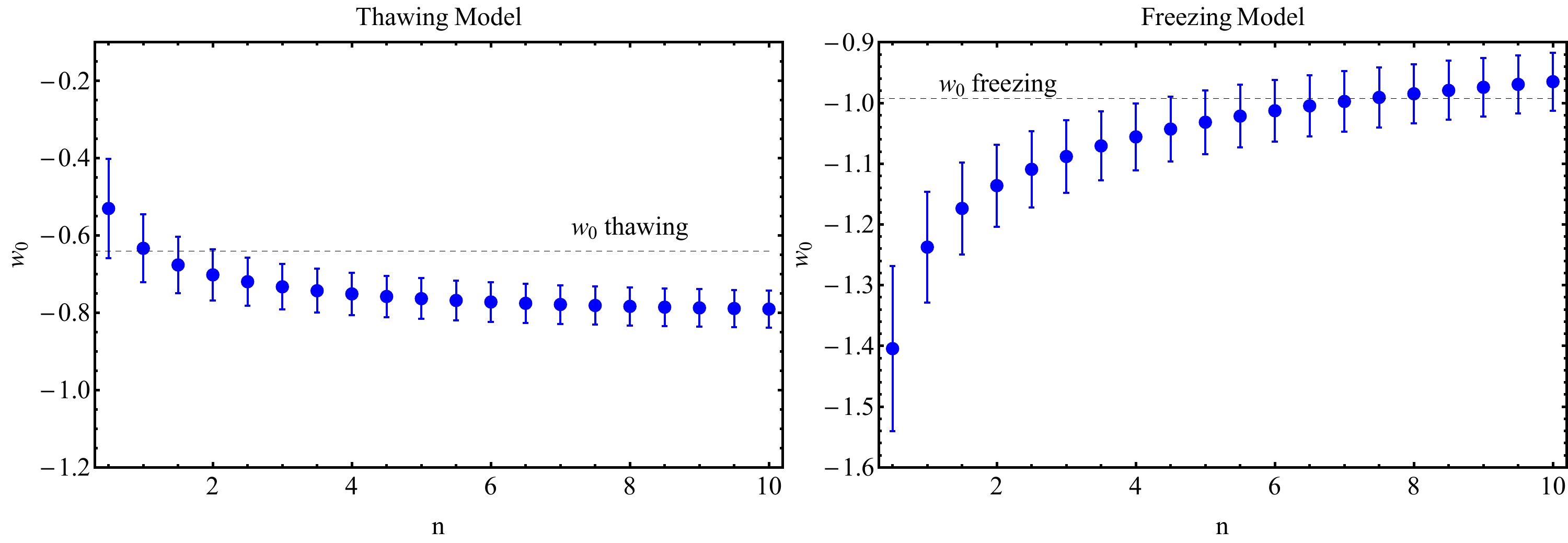}}}
\caption{For a thawing (left) and freezing (right) underlying model, the figures shows best fit values of $w_0$ as a function of $n$ with $1\sigma$ errors, as well as the true value of $w(z=0)$ in the context of each model (dashed lines).}
\label{fig:Figure_15}
\end{figure*}

\begin{figure}[!th]
\centering
\vspace{0cm}\rotatebox{0}{\vspace{0cm}\hspace{0cm}\resizebox{0.49\textwidth}{!}{\includegraphics{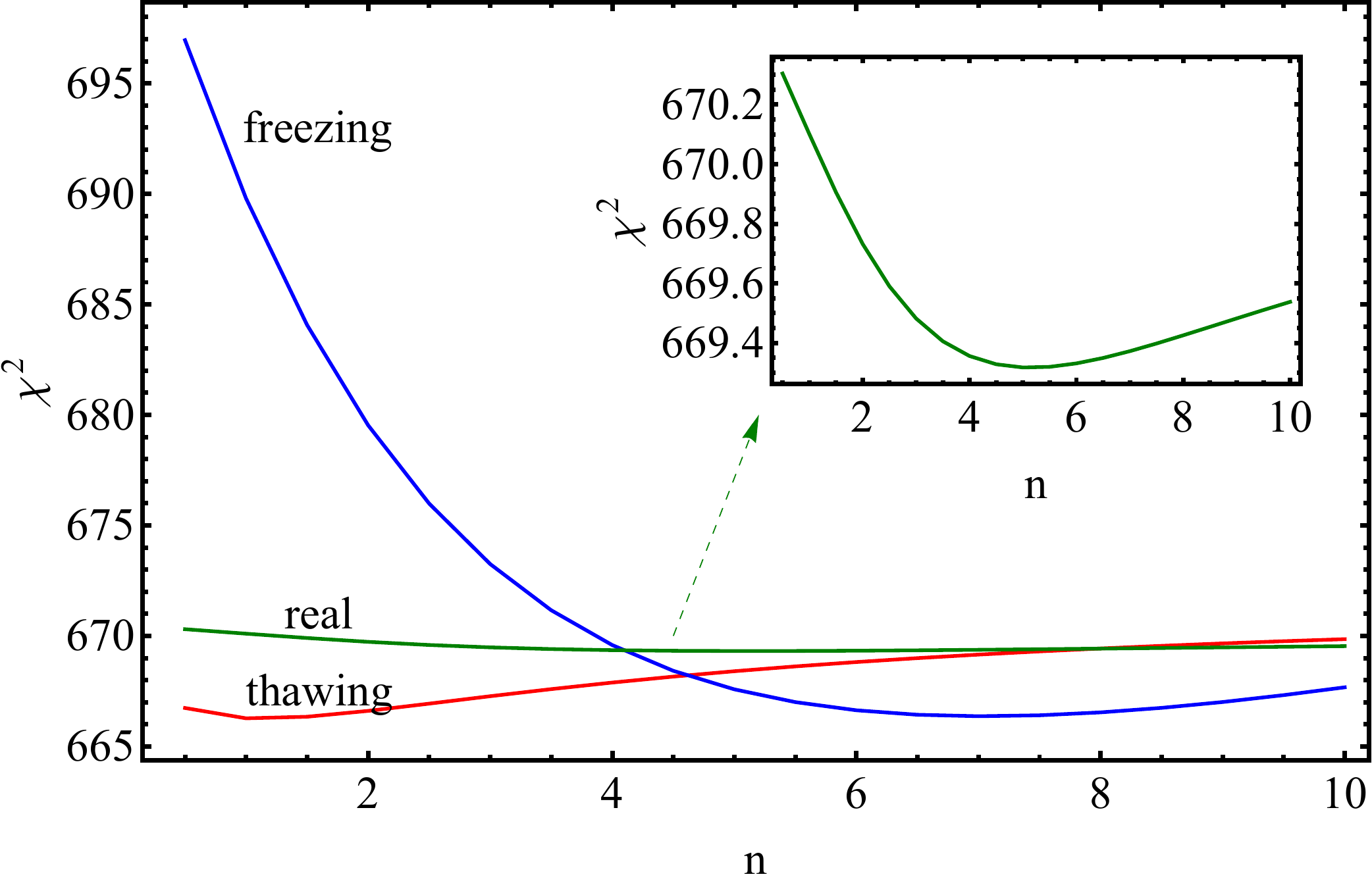}}}
\caption{The $\chi^2$ as a function of $n$ for a thawing underlying model (red line), a freezing underlying model (blue line) and for real data (green line) using Fig.~\ref{fig:Figure_13}. The $\chi^2$ for \lcdm \ for a thawing underlying model, a freezing underlying model and the real data are $686.534, 734.314$ and $678.537$ respectively. Notice that the red line has a minimum for $n=1$ (CPL), the blue one for $n=7$ (7CPL), while the green one is slightly preferable for $n=5$.}
\label{fig:Figure_16}
\end{figure}

Recent non-parametric constraints \cite{Wang:2013zfa, Qi:2009yr} on the dark energy equation of state derived from a WMAP+SNLS+BAO+H(z) dataset \cite{Said:2013jxa} are shown in Fig.~\ref{fig:Figure_13}.

As shown in Fig.~\ref{fig:Figure_13} the constraints on $w(z)$ are significantly more stringent for low redshifts $z$ where there are more data available. Since thawing models predict more deviation from $w=-1$ at low $z$, these models are strongly constrained to be close to \lcdm. In contrast, freezing models deviate from $w=-1$ at high $z$ where the constraints are weaker. Therefore this class of models is less constrained and is allowed to deviate more from \lcdm.

Using the errorbars in the redshift bins shown in Fig.~\ref{fig:Figure_13} we may generate scattered Gaussian mock data in each redshift bin (we used 50 points in each bin) with mean equal to the predicted value provided by a quintessence model (thawing or freezing) and standard deviation equal to the one indicated by the observational constraints of Fig.~\ref{fig:Figure_13}.

These data are shown in Fig.~\ref{fig:Figure_14} for a thawing underlying model (left) and a freezing underlying model (right) with  parameters selected so that there is consistency with observational constraints. In the same figure we show the numerically obtained $w(z)$ of the underlying quintessence cosmological model and two best fit parametrizations in each case (one thawing (CPL) and one freezing (7CPL)). The fit was made using least squares on the scattered Gaussian data points. Notice that the best fit parametrization is close to the underlying model $w(z)$ only in the case when the parametrization is in the same class as the underlying quintessence model.

We next address the following question: How does the estimated  value of present day equation of state parameter $w(z=0)$ depend on the type of parametrization used to fit the $w(z)$ data?

In order to answer this question we fit the simulated data of Fig.~\ref{fig:Figure_14}a (thawing) and Fig.~\ref{fig:Figure_14}b (freezing) using the nCPL parametrization of equation \eqref{ncpl} for different values of $n$. We have seen that for $n=1$ the parametrization reduces to the usual CPL which belongs to the thawing class while for larger $n$ the parametrization transforms to one in the freezing class.  The best fit parameter value $w_0$ is identical to the predicted value for $w(z=0)$ in the context of the nCPL parametrization. Thus, in Fig.~\ref{fig:Figure_15} we plot the best fit value of $w_0$ (with $1\sigma$ errors) as a function of $n$ for an underlying thawing (left) and for an underlying freezing model (right). The true underlying value of $w(z=0)$ is also indicated in each plot by a dashed line for comparison with the estimated value of $w_0$ as a function of $n$. For a thawing underlying model (Fig.~\ref{fig:Figure_15}a) the standard CPL ($n=1$) provides an excellent estimate for the value of $w_0=w(z=0)$ while higher values of $n$ are less accurate. In contrast for a freezing underlying model (Fig.~ \ref{fig:Figure_15}b) the standard CPL provides a poor estimate for the value of $w(z=0)$. For $n=1$, the true value is more than $2\sigma$ away from the estimated value while for $n=7$ (freezing parametrization) the estimated value of $w_0$ is in excellent agreement with the true value. We conclude that the use of a parametrization that is incompatible with the underlying model can lead to significant misleading conclusions regarding the present value of the equation of state parameter.

A guide for the selection of the right parametrization may be provided by using the likelihood for comparison of candidate parametrizations to be used with a given set of $w(z)$ data. In Fig.~\ref{fig:Figure_16} we show a plot of the log-likelihood, i.e. the $\chi^2$, as a function of $n$ corresponding to the quality of fit of nCPL to simulated data originating from a freezing underlying model (blue line) and from a thawing underlying model (red line). For the case of a freezing underlying model there is a minimum of the $\chi^2$ at $n=7$ (preferred value) while in the case of a thawing underlying model the minimum of the $\chi^2$ is at $n=1$ as expected. Therefore, the $\chi^2$ may be used as a criterion for the selection of the suitable parametrization class corresponding to a given dataset.

The simulated data used in this section are extensive in redshift space and useful for testing our arguments but they do not correspond directly to particular cosmological observations. Instead they emerge as combined constraints emerging from several datasets. Thus the simulated scattered points do not correspond directly to particular data points coming from a given cosmological dataset.

It would be desirable to test the two classes of parametrizations we have proposed using simulations of more realistic datapoints like Type Ia supernovae. Such data points are more concentrated towards low redshifts where the properties of the two parametrization classes may not be as distinct. However, it is important to compare  parametrizations in the two classes using directly simulated realistic data. Thus we focus on this approach in the next section.

\section{Simulated datasets at the level of luminosity distance $d_L(z)$}
\label{sec:Section 4}

In order to compare the two classes of parametrizations in more realistic data we construct simulated datasets based on the {\it{Union 2.1}} compilation. This compilation consists of 580 datapoints expressed as distance moduli in a redshift range $z\in [0,1.414]$.

There are three main significant new features of this compilation compared to the data used in the previous section:
\begin{itemize}
\item[1.] They are more realistic datapoints based on real observations (SnIa) with error bars and without the need for binning. Thus, in this case we do not introduce a larger number of scattered points to emulate a small number of binned data points with errorbars. The number of the simulated data points is the same as the number of the original points and they have the same errorbars. Their mean value however, is determined by the simulated quintessence model.

\item[2.] Since these points originate from a single type of cosmological observation they have a more limited redshift range. They are more constraining at low redshifts (up to and $z\sim 0.8$) and they include no information from higher redshift probes (\eg CMB), as were the data used in the previous section.

\item[3.] The data are presented at the level of the luminosity distance $d_L(z)$, which is a directly measurable quantity, and not at the level of the equation of state $w(z)$, which can only be inferred in practice under certain assumptions, \eg knowledge of $\Omega_m$ \etc Thus we have to convert the $w$ parametrizations $w(z,w_0,w_a)$ to parametrizations of the form $d_L(z,w_0,w_a)$ and for that we need to calculate the dark energy density $\rho_\text{DE}(z)$.
\end{itemize}

For a general equation of state $w(z)$ the dark energy density $\rho_\text{DE}(z)$ is given by
\be
\rho_\text{DE}(a)=\rho_\text{DE}(a=1)~e^{-3\int_1^a \frac{1+w(a^\prime)}{a^\prime}da^\prime},
\ee
so that for the nCPL model of equation \eqref{ncpl} we have
\ba
\Omega_\text{DE}(a)&=&\Omega_{\text{DE},0}~a^{-3(1+w_0+w_a)}\cdot\nn \\ &&e^{-3w_a\left(H_n-a~n~{}_3F_2\left(1,1,1-n;2,2;a\right)\right)},
\ea
where $H_n\equiv\sum^n_{i=1}1/i$ is the n-th {\it harmonic number} and ${}_3F_2(1,1,1-n;2,2;a)$ is a hypergeometric series. Then, the Hubble parameter for a flat universe containing only matter and dark energy is given by
\ba
H(z)^2/H_0^2&=&\Omega_{m,0} a^{-3}+(1-\Omega_{m,0})~a^{-3(1+w_0+w_a)}\nn \\
&\cdot&e^{-3w_a\left(H_n-a~n~{}_3F_2\left(1,1,1-n;2,2;a\right)\right)}.
\ea

\begin{figure}[!t]
\centering
\vspace{0cm}\rotatebox{0}{\vspace{0cm}\hspace{0cm}\resizebox{0.49\textwidth}{!}{\includegraphics{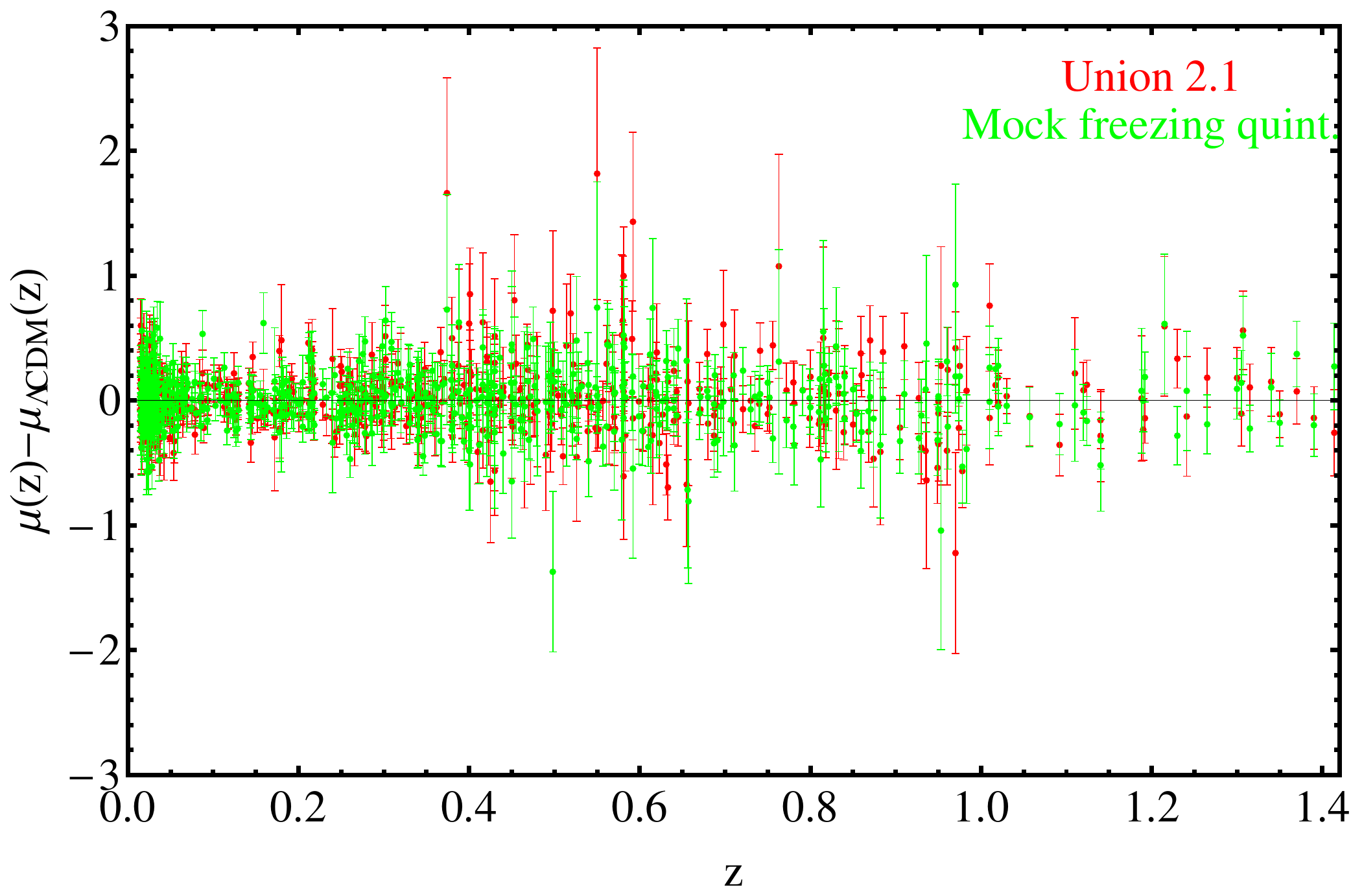}}}
\caption{Simulated mock data corresponding to an underlying freezing model (green points) along with the actual Union2.1 dataset (red points).}
\label{fig:Figure_17}
\end{figure}

With these in mind, we generate the simulated data points by replacing each real distance modulus data by a point with the same errorbars selected randomly using a Gaussian probability distribution with standard deviation equal to the $1\sigma$ error of the point and mean equal to the value predicted by the quintessence model to be simulated. In Fig.~\ref{fig:Figure_17} we show a simulated dataset corresponding to a freezing quintessence model (green points) along with the actual Union 2.1 dataset (red points).

The theoretical distance moduli that are to be compared with the observed distance moduli of the Union 2.1 compilation are written as \be \mu_\text{th}(z_i)\equiv 5\log_{10} D_L(z_i)+\mu_0, \label{th_dist_mod} \ee
where $\mu_0\equiv 42.38-5\log_{10}h$, and in a flat universe the Hubble-free luminosity distance $D_L=H_0 d_L$ ($d_L$ denotes the physical luminosity distance) is
\be
D_L(z)=(1+z) \int_0^z \frac{dz^\prime}{E\left(z^\prime;\Omega_{m,0}, w_0, w_a\right)}, \label{lum_dist}
\ee
where $E\left(z;\Omega_{m,0}, w_0, w_a\right)\equiv H\left(z;\Omega_{m,0}, w_0, w_a\right)/H_0$. Finally, the $\chi^2$ for the SnIa data can be calculated as \be \chi_\text{Sn}^2(w_0, w_a)=\sum_{i=1}^{580} \frac{\left[\mu_\text{obs}(z_i)-\mu_\text{th}(z_i;w_0,w_a)\right]^2}{\sigma_i^2}. \label{chi_sq} \ee

\begin{figure*}[!t]
\centering
\vspace{0cm}\rotatebox{0}{\vspace{0cm}\hspace{0cm}\resizebox{0.49\textwidth}{!}{\includegraphics{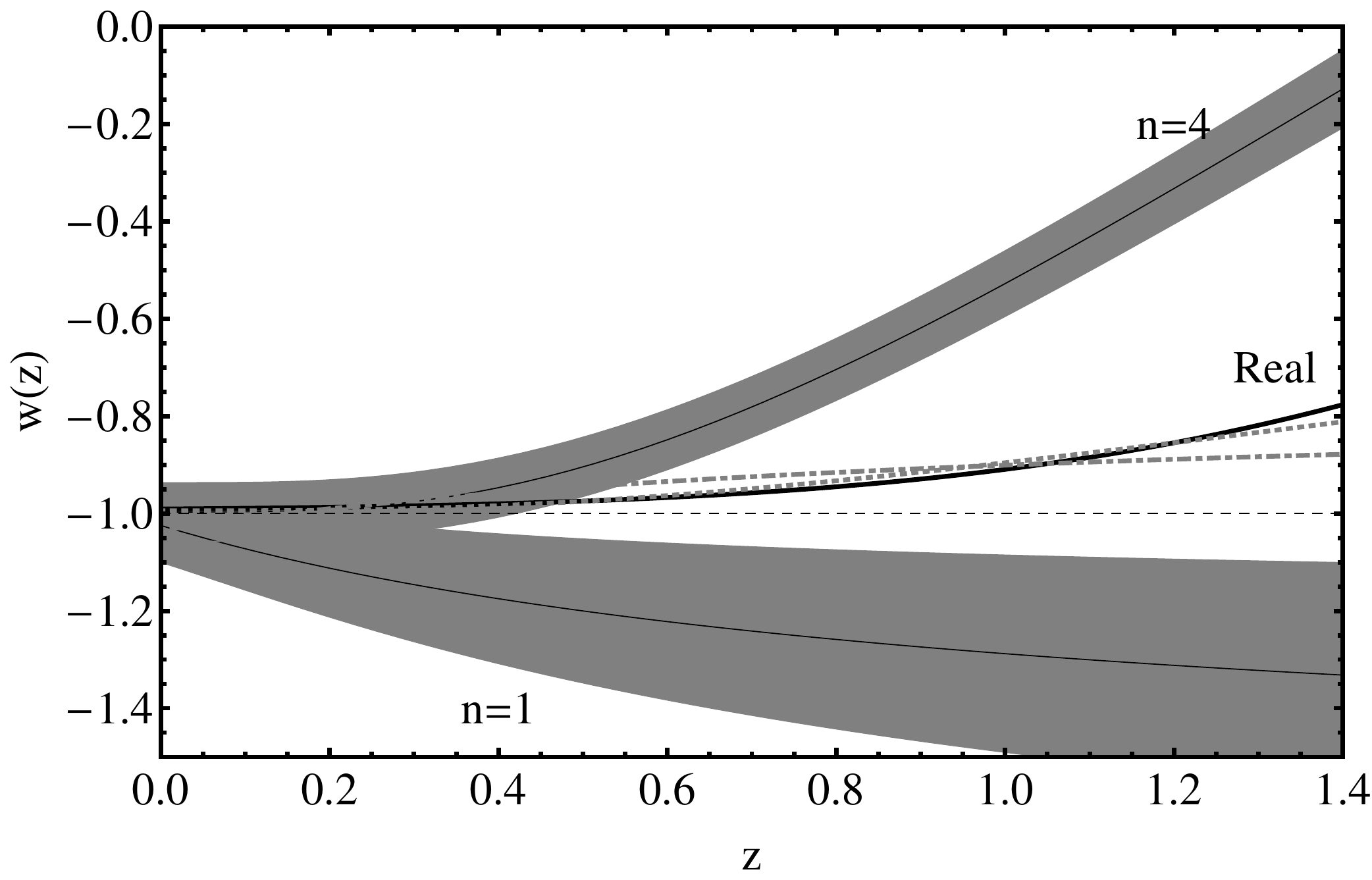}}}
\vspace{0cm}\rotatebox{0}{\vspace{0cm}\hspace{0cm}\resizebox{0.49\textwidth}{!}{\includegraphics{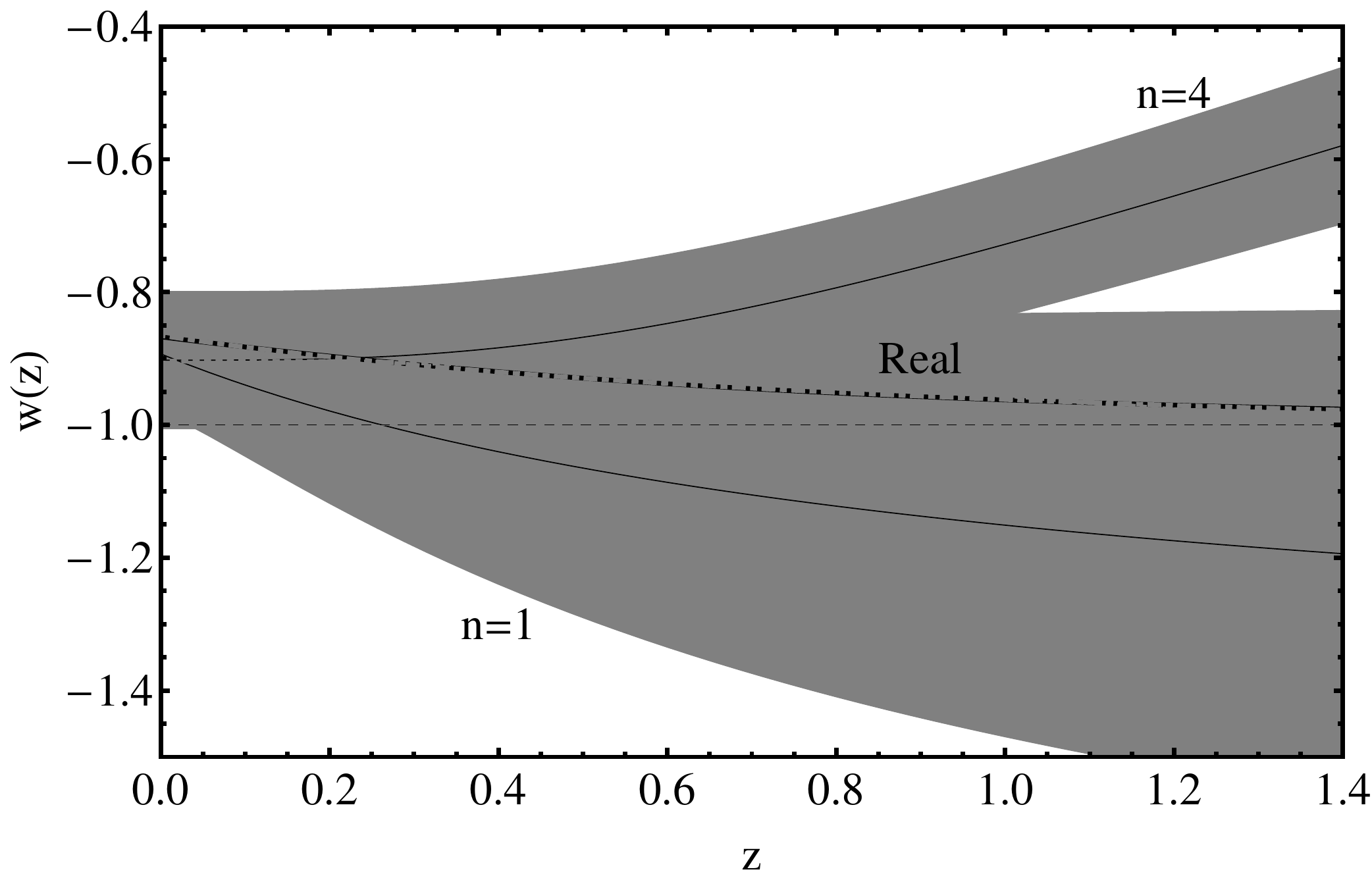}}}
\caption{Left: The $w(z)$ of the freezing underlying model superimposed with the best fit $w(z)$ for $n=1$ and $n=4$ (solid lines with gray regions) using the simulated Union 2.1 data. We also show the best fit obtained directly with least squares from the underlying curve $w(z)$ using again nCPL with $n=1$ and $n=4$ (dashed and dotted lines, no simulated data in this case).  Right: The $w(z)$ of the thawing underlying model superimposed with the best fit $w(z)$ for $n=1$ and $n=4$ (solid lines with gray regions) using the simulated Union 2.1 data. We also show the best fit obtained directly with least squares from the underlying curve $w(z)$ using again nCPL with $n=1$ and $n=4$ (dashed and dotted lines, no simulated data in this case). \label{fig:Figure_18}}
\end{figure*}

We minimize the $\chi^2$ using the simulated Union 2.1 data produced with a freezing or thawing underlying model using the nCPL parametrizations with fixed $n$ after also marginalizing analytically over $\mu_0$ and numerically over $\Omega_m$ \cite{Nesseris:2005ur}.

\begin{figure*}[!t]
\centering
\vspace{0cm}\rotatebox{0}{\vspace{0cm}\hspace{0cm}\resizebox{0.49\textwidth}{!}{\includegraphics{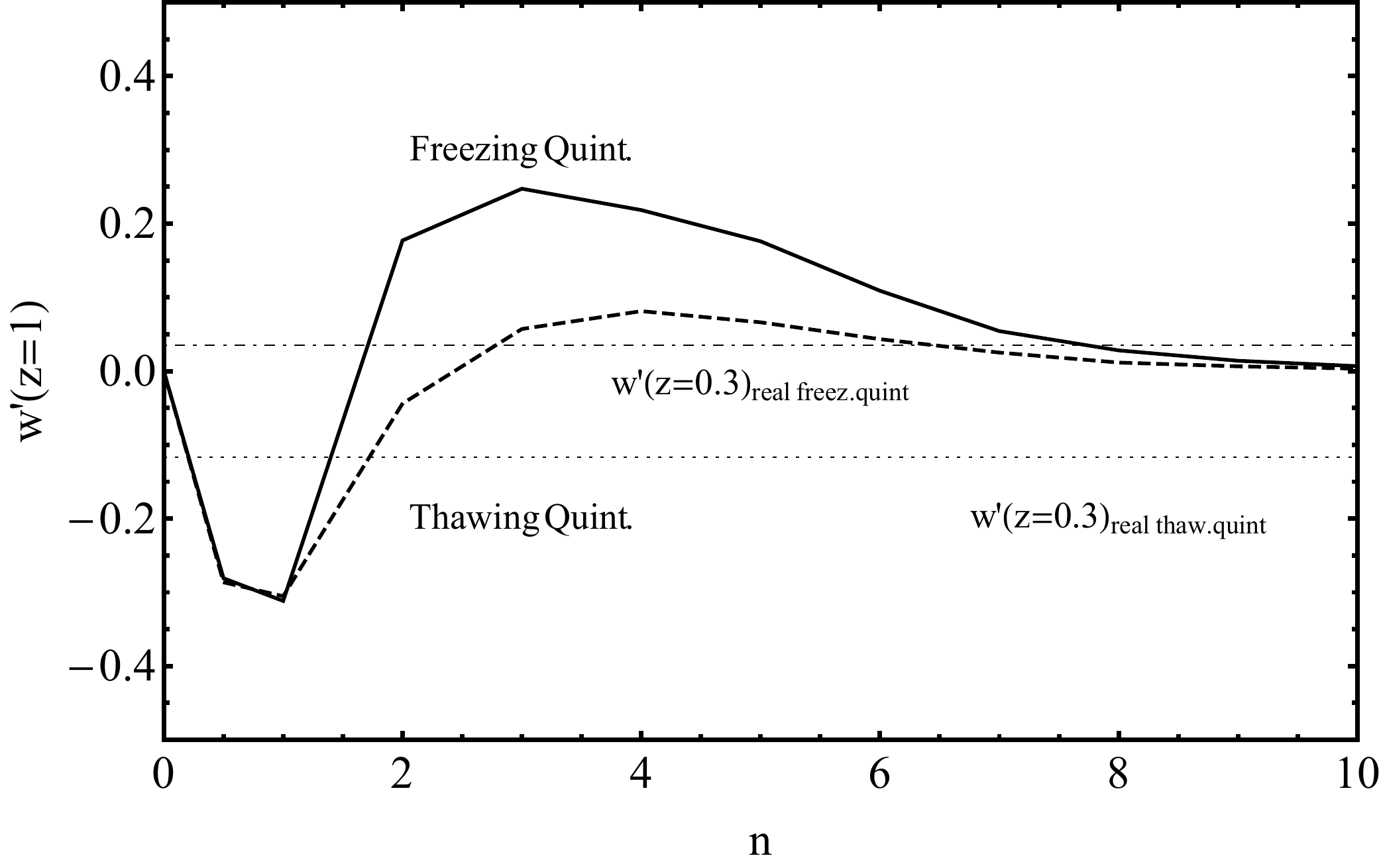}}}
\vspace{0cm}\rotatebox{0}{\vspace{0cm}\hspace{0cm}\resizebox{0.49\textwidth}{!}{\includegraphics{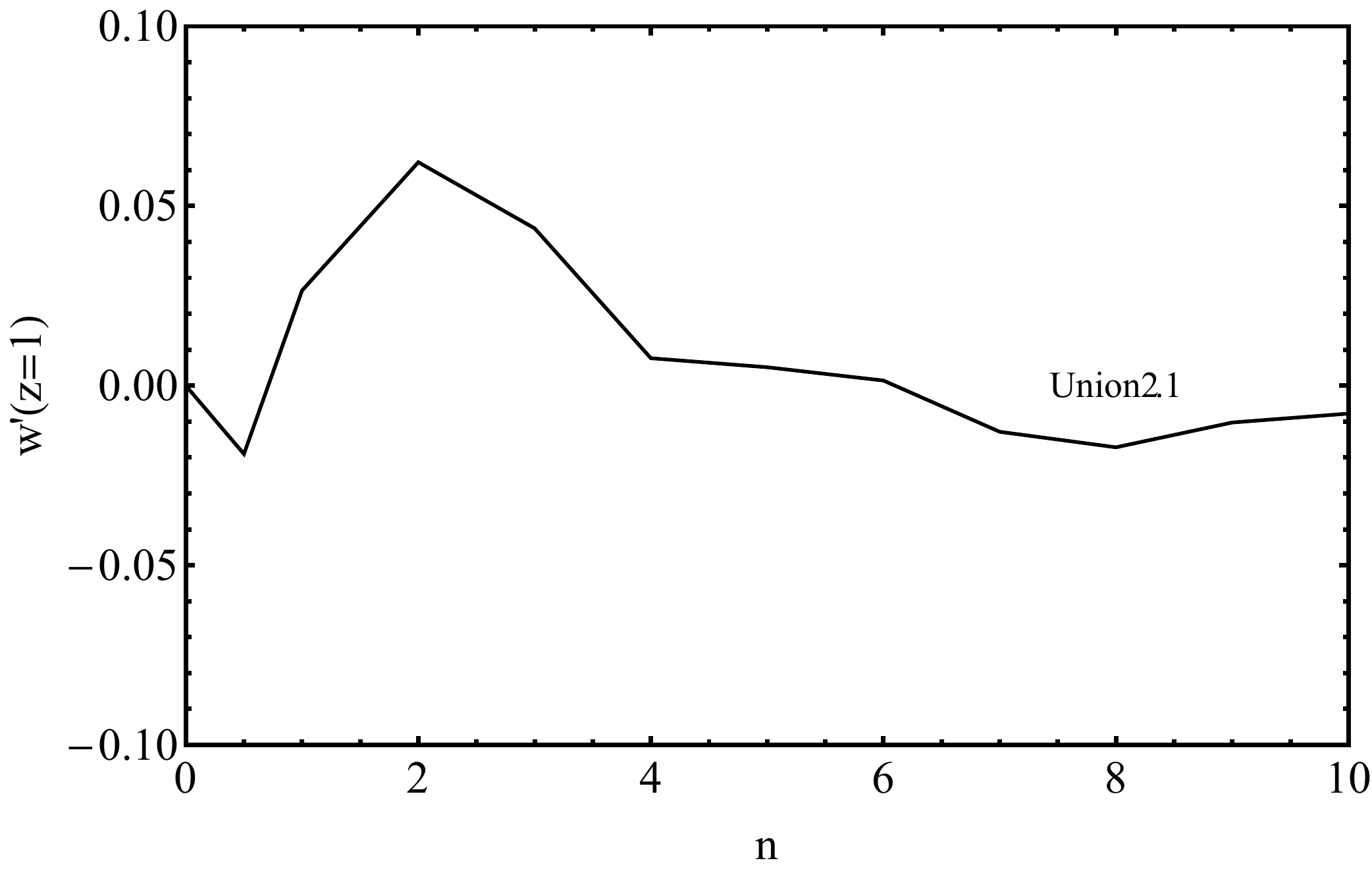}}}
\caption{Left: The slope of the best fit $w(z)$ for the freezing (solid line) or thawing (dashed) models at $z=0.3$, as a function of $n$. The real slopes  for the freezing and thawing models are indicated by the dot-dashed and dotted lines respectively. Right: The slope of the best fit $w(z)$ at $z=0.3$ as a function of $n$ for the real Union 2.1 data.\label{fig:Figure_19}}
\end{figure*}

In Fig.~\ref{fig:Figure_18}a (left) we show $w(z)$ of the freezing underlying model superimposed with the best fit $w(z)$ for $n=1$ and $n=4$ (solid lines with gray regions) using the simulated Union 2.1 data. We also show the best fit obtained directly with least squares from the underlying curve $w(z)$ using again nCPL with $n=1$ and $n=4$ (dashed and dotted lines, no simulated data in this case). Notice that for the direct fit to the underlying model the value of the predicted value of $w(z=0)$ is significantly smaller than the true value. On the other hand the slope of the fit $w(z)$ is roughly consistent with the true slope for intermediate redshifts ($z>0.1$). Thus, it this case where there is uniform information over all redshifts probed, the use of a unsuitable parametrization ($n=1$) leads to an incorrect value of $w_0$ but approximately correct value of the slope at intermediate $z$.

When the simulated Union 2.1 data are used for the fit the predicted value of $w_0$ comes much closer to the true value (see Fig.~\ref{fig:Figure_18}a). This result is obtained  because the SnIa data are more constraining at low redshifts and any deviation from the true value of $w_0$ is translated in significant increase of $\chi^2$. In this case, the error induced by using the unsuitable parametrization is transferred to the predicted slope of $w(z)$ at relatively  high redshift ($z\in [0.2,1.2]$) which is significantly different from the true slope of $w(z)$. Indeed for $n=1$ the best fit $w(z)$ is a decreasing function of $z$ while the true underlying $w(z)$ is a rapidly increasing function (at relatively high $z$).

On the contrary, when a suitable (freezing) parametrization is used ($n=4$), both the slope and the value of $w(z=0)=w_0$ are in agreement with the underlying freezing model as demonstrated in Fig.~\ref{fig:Figure_18}a (left).

Corresponding conclusions for the case of a thawing underlying model can be drawn by inspecting Fig.~ \ref{fig:Figure_18}b (right). In this case the suitable parametrization is thawing with $n=1$ which provides good agreement for both the slope and the present day value $w(z=0)=w_0$. On the other hand, a freezing parametrization ($n=4$) provides approximately correct value for $w(z=0)=w_0$ (due to the increased density of datapoints at low $z$) but an incorrect value of the slope of $w(z)$ at high $z$.

Therefore, when cosmological data are more constraining at low redshifts, as is the case for the Union 2.1 data, the basic misleading effect of using an inappropriate parametrization to fit the data is not an incorrect value for $w_0$, but an incorrect value of the slope of $w(z)$ at relatively intermediate or high redshifts. To demonstrate this fact, we show in Fig.~\ref{fig:Figure_19}a (left) a plot of the slope of the best fit $w(z)$ over the true slope of the underlying (freezing or thawing model) at $z=0.3$, as a function of $n$. Clearly, the best fit slope deviates significantly from the true slope when an unsuitable parametrization is used ($n\neq 1$ for thawing and $n<4$ for freezing underlying model). The best fit slope of the actual Union 2.1 data as a function of $n$ is shown in Fig.~\ref{fig:Figure_19}b (right).

\section{Conclusions-Discussion}
\label{sec:Conclusion}

We have demonstrated that the dark energy parametrizations fitting the equation of state parameter $w(z)$  may be divided in two classes depending on their convexity properties. Based on these properties we have divided these parametrizations in two classes: {\it{thawing parametrizations}} and {\it{freezing parametrizations}}. Each class is suitable for fitting the corresponding class of quintessence cosmological models or more general models with the same convexity. We have proposed a three parameter parametrization (nCPL) that interpolates between the two classes of parametrizations by varying one of its parameters ($n$) and reduces to the usual CPL parametrization in its thawing limit ($n=1$).

We have also shown that the correct class of parametrizations  for a given cosmological dataset may be identified using the nCPL parametrization for various values of $n$ and ploting the log-likelihood as a function of $n$. The value of $n$ that minimizes the $\chi^2$ can determine the correct class of parametrizations to be used in the fit and identifies the most probable class of the underlying cosmological model.

Alternatively, a non-parametric method \cite{Huterer:2004ch, Zhao:2012aw, Zhao:2015wqa, Holsclaw:2010sk, Sullivan:2007pd, Ruiz:2012rc, Shafieloo:2005nd, Nesseris:2014vra} (like principal component analysis or null tests \cite{Shafieloo:2012rs, Sahni:2008xx}) may be used to reconstruct $w(z)$ using independent redshift bins assuming that $w(z)$ is constant in each bin. In this case however it is more difficult to extract information about the slope of $w(z)$ and future trends for its evolution.

Finally we have pointed out the misleading conclusions that can be obtained if an unsuitable parametrization is used to fit a particular dataset. Such misleading conclusions depend on the distribution of datapoints in redshift space. If the dataset is more constraining at low redshifts (like the Union 2.1 dataset)  then the error introduced on the estimated present day value of $w(z)$ is relatively small and depends weakly on the class of parametrization used. However, in this case the use of unsuitable parametrization estimate for the slope of $w(z)$ at relatively intermediate and leads to incorrect high redshifts. This error is drastically reduced with the use of a parametrization of suitable class.

On the other hand if a dataset is relatively uniform in redshift space, then the slope of the best fit $w(z)$ depends weakly on the class used but the estimated present value of $w(z)$ can be more than $2\sigma$ away from the true value if an unsuitable parametrization is selected.

The Union 2.1 dataset has been shown to have a better overall agreement of the slope $w^\prime(z=0.3)$ with the freezing quintessence mock data, thus indicating a mild preference for high values of $n$ for the nCPL parametrizations but also for the freezing quintessence models. However the value of  the $\chi^2$ for \lcdm \ remains significantly lower at $564.284$. This preference for a high $n$ nCPL parametrization and the freezing quintessence models will become more clear in the future as the high redshift data increase and the dataset becomes more uniform at high $z$.
\\ \\
\textbf{Numerical Analysis Files}: See Supplemental Material at \href{https://www.dropbox.com/s/0gyal92olbcwabt/parametrizations.zip?dl=0}{here} for the mathematica files used for the production of the figures, as well as the figures.

\section*{Acknowledgements}
S.N. acknowledges support from the Research Project of the Spanish MINECO, FPA2013-47986-03-3P, the Centro de Excelencia Severo Ochoa Program SEV-2012-0249 and the Ram\'{o}n y Cajal programme through the grant RYC-2014-15843.

\raggedleft
\bibliography{bibliography}

\end{document}